\documentclass[a4paper,twocolumn,showpacs,superscriptaddress,amsmath,amssymb,aps,pra,preprintnumbers]{revtex4-1}
\usepackage{bbm}
\usepackage{amsmath}
\usepackage{verbatim}
\usepackage{graphicx}
\usepackage{braket,amsmath}
\usepackage[dvipsnames, svgnames, x11names]{xcolor}
\usepackage{array}
\usepackage{bibunits}
\usepackage{braket}
\usepackage{mathrsfs}
\usepackage{amsmath}
\usepackage{graphicx}% Include figure files
\usepackage{dcolumn}% Align table columns on decimal point
\usepackage{bm}% bold math
\usepackage[capitalise]{cleveref}
\crefname{section}{Sec.}{Secs.}
\crefname{appendix}{App.}{Apps.}
\begin{document}

%\preprint{APS/123-QED}

\title{Entangling spins using cubic nonlinear dynamics}% Force line breaks with \\
%\thanks{A footnote to the article title}

\author{Lingxia Wang}
 \affiliation{%
 Department of Physics, Wenzhou University, Zhejiang 325035, China
}
 \author{Yani Wang}%
%\email{liugang@stu.wzu.edu.cn}
\affiliation{%
 Department of Physics, Wenzhou University, Zhejiang 325035, China
}%
 \author{Yujing Cheng}%
%\email{liugang@stu.wzu.edu.cn}
\affiliation{%
 Department of Physics, Wenzhou University, Zhejiang 325035, China
}
\author{Zhiqi Yan}%
%\email{liugang@stu.wzu.edu.cn}
\affiliation{%
	Department of Physics, Wenzhou University, Zhejiang 325035, China
}
\author{Lei Xie}%
%\email{liugang@stu.wzu.edu.cn}
\affiliation{%
	Department of Physics, Wenzhou University, Zhejiang 325035, China
}
\author{Gang Liu}%
%\email{liugang@stu.wzu.edu.cn}
\affiliation{%
	School of Physical Science and Technology, Lanzhou University, Lanzhou 730000, China}
 \author{\\Jinmin Fan}%
%\email{liugang@stu.wzu.edu.cn}
\affiliation{%
 Department of Physics, Wenzhou University, Zhejiang 325035, China
}
 \author{Di Wang}%
%\email{liugang@stu.wzu.edu.cn}
\affiliation{%
 Department of Physics, Wenzhou University, Zhejiang 325035, China
}
\author{Yiling Song}%
%\email{liugang@stu.wzu.edu.cn}
\affiliation{%
 Department of Physics, Wenzhou University, Zhejiang 325035, China
}
\author{Linli He}%
\email{linlihe@wzu.edu.cn}
\affiliation{%
 Department of Physics, Wenzhou University, Zhejiang 325035, China
}
 \author{Wei Xiong}%
\email{xiongweiphys@hotmail.com}
\affiliation{%
 Department of Physics, Wenzhou University, Zhejiang 325035, China
}%
\author{ Mingfeng Wang}%
 \email{mfwang@wzu.edu.cn}
\affiliation{%
 Department of Physics, Wenzhou University, Zhejiang 325035, China
}%



%%%%%%%%%%%%%%%%%%%%%%%%%%%%%%%%%%%%%%%%%%%%%%%%%%%%%%%%%%%%%%%%%%%%%%%%%%%%%%%%%%%%%%%%%%%%%%%%%%%%%%

\begin{abstract}
Entangled states with a large number of $N$ atomic spins are a key ingredient for quantum information processing and quantum metrology. Nowadays, the preparation of such states has mainly relied on the quadratic nonlinear dynamics. Here, we investigate the preparation of spin-spin multipartite entanglement, witnessed by quantum Fisher information, by using the cubic nonlinear dynamics. We find that, in the regime of weak coupling,  the cubic scheme can greatly speed up the rate of entanglement generation as compared to the quadratic scheme (about $N$ times faster). In the strong coupling regime, the cubic nonlinear dynamics enables
the periodic in time generation of a broad variety of new-type macroscopic superposition states, which allow us to realize near-Heisenberg-limit phase sensitivity.  In addition, we also reveal an interesting feature that the amount of entanglement generated by the cubic scheme has a macroscopic sensitivity to the parity of $N$, which has no counterpart in quadratic nonlinear dynamics and can be exploited for sensing the parity of $N$ at the single-spin level. We also propose a new approach for a fast and high-fidelity generation of maximally entangled Greenberger-Horne-Zeilinger (GHZ) states. By using an alternative cubic-quadratic-admixture type of nonlinear interaction, we show that one may accelerate the procedure of GHZ-state generation. The realization of the cubic nonlinear dynamics is also considered, showing that the cubic nonlinear dynamics can be realized by either repeatedly using linear- and quadratic-nonlinear dynamics or utilizing light-mediated interactions in just one step. Finally, by taking realistic imperfections into account, we find that the cubic scheme is sensitivity to the single-spin decay in the strong coupling regime, while is robust against the collective dephasing. Our proposed schemes offer potential possibilities for realizing
high-sensitivity metrology in a variety of platforms, including trapped ions and cold or warm atomic ensembles.
%\begin{description}
%\end{description}
\end{abstract}

\maketitle

%\tableofcontents
\section{INTRODUCTION }
%\section{\label{sec:level1}First-level heading:\protect\\ The line
%break was forced \lowercase{via} \textbackslash\textbackslash}
The generation of entanglement between a large number of spins is an extremely important subject in precision metrology and quantum science. In quantum metrology \cite{PhysRevLett.96.010401}, highly entangled spin states enable precision metrology beyond the standard quantum limit (SQL) \cite{PAS.106}, even approaching the Heisenberg limit (HL) \cite{PhysRevA.65.053819,PhysRevA.55.2598}. In the field of quantum information \cite{RevModPhys.77.513,RevModPhys.81.1727}, entangled spin ensembles are not only recognized as key resources for quantum communication \cite{PhysRevLett.85.5643,Nature.413.6854,PhysRevA.79.012327} but also considered as a promising platform for quantum computation \cite{Book1,2021Linear,Barrett_2010,PhysRevA.83.062339}.

To date, a variety of approaches have been developed for producing entangled states of spin ensemble, which can be classified into two main categories. One is based on the projection measurement (such as quantum nondemolition measurement) \cite{Kuzmich1998,PhysRevLett.86.4431,PhysRevLett.102.033601,Nature.581.7807}: first, entanglement is established between the spin system and an auxiliary quantum system (usually a light field), and then, a measurement of the auxiliary quantum system will project the spin state into a multipartite entangled state. Another one has relied on unitary evolution of an initial product spin state under a nonlinear spin-spin (NSS) interaction. Among these NSS interactions, the most widely studied one is possible the one-axis-twisting (OAT) interaction  \cite{PhysRevA.47.5138,PhysRevLett.104.073602,riedel2010atom,gross2010nonlinear,hosten2016quantum,PhysRevA.99.043840}, which, as shown by Ueda \emph{et al.} \cite{PhysRevA.47.5138}, can produce pairwise spin-spin entanglement that is the origin of spin squeezing \cite{PhysRevA.68.012101,ma2011quantum}. Spin squeezing is probably the most sought-after multipartite entangled resource in the field of quantum metrology, as the phase estimation based on spin squeezing is comparatively easy to implement in realistic experiments \cite{ma2011quantum}. Up till now, most studies of NSS interaction have mainly been concentrated on how to efficiently create highly squeezed spin states, such as two-axis-twisting interaction \cite{cappellaro2009quantum,PhysRevLett.107.013601,borregaard2017one,groszkowski2020heisenberg,PhysRevA.96.013823}, twist-and-turn interaction \cite{PhysRevA.63.055601,PhysRevA.66.043621,PhysRevA.92.023603,PhysRevLett.107.013601}, and twisting-tensor interaction \cite{PhysRevA.91.053826}. However, squeezed spin states are only one category of multipartite entangled states that can benefit the quantum metrology. Other categories of entangled states, although having no spin-squeezing property, may also be useful for quantum metrology and sensing, such as the GHZ state enabling the phase sensitivity reaching the HL \cite{RevModPhys.90.035005}. In fact, apart from the spin squeezing, the quantum Fisher information (QFI) provides
a more general and profound way to estimate whether a given spin state is useful or not for quantum metrology \cite{PhysRevLett.102.100401,PhysRevA.85.022321,PhysRevA.85.022322,Science345.424,Phys7406,PhysRevA.92.012312}. The larger the
QFI of the entangled state, the more useful the state might be. Therefore, it is of particulary necessary to reconsider the entanglement generation induced by the NSS interactions from the perspective of QFI. Accordingly, discovering and devising new NSS-interaction schemes that can rapidly and efficiently generate large QFI is of vital importance for realizing high-sensitivity metrology.

In this paper, we propose to use \emph{cubic} NSS interaction to entangle individual spins. Although exhibiting no spin squeezing, the entangled state created by the cubic interaction have several advantages over the quadratic interaction from the perspective of QFI. First, in the weak-coupling regime we find that the cubic scheme can produce QFI (and thus entanglement) much more rapidly than the quadratic one. Quantitative analysis indicates that the acceleration rate is proportional to the spin number of the system. The cubic scheme thus offers a great advantage over the quadratic scheme in the case of large spin systems. Second, the QFI of the cubic scheme in the strong-coupling regime oscillates fast with coupling strength, which, on average, is larger than the QFI produced by the quadratic scheme. Besides, the cubic NSS interaction enables the production of a broad variety of macroscopic superposition states that have  large QFI, which has no counterpart in the quadratic NSS dynamic.

We also analyze an interesting phenomena that has not yet been discovered previously. That is, the QFI production of the cubic scheme is extremely sensitivity to the parity of the total spin number $N$. We find that the amount of QFI at a specific instant of time for even $N$ spins versus odd $N+1$ spins change dramatically from $N$ (corresponding to no entanglement among spins) to $N^2$ (maximal entanglement). This \emph{entanglement} even-odd effect is quite different from the one exhibited by the OAT interaction \cite{PhysRevA.56.2249}, which, as we will show later, is an orientation even-odd effect. We also show that this entanglement even-odd effect enables us to design a new type of sensing modality to detect the parity of the total spin number of a spin system at the single-spin level.

Apart from the cubic NSS interaction, we also have studied a hybrid NSS interaction---cubic-quadratic-admixture (CQA) interaction, which is a weighted sum of the cubic and the quadratic interaction. We find that the CQA interaction is an excellent tool for preparing the GHZ states. High-fidelity GHZ states could be created by simply applying the CQA evolution to the spin system for a certain time interval. In contrast to the OAT scheme \cite{PhysRevA.56.2249}, our hybrid scheme can greatly accelerate the procedure of GHZ-state generation, which tremendously eases experimental requirements.

To realize the cubic interactions in realistic spin systems, two approaches have been developed. One utilizes the linear and quadratic interactions. Unlike the harmonic oscillator systems, where high-order interactions can not be constructed from the quadratic interactions (known as the Gaussian operations) \cite{PhysRevLett.82.1784,PhysRevLett.97.110501}, we show that the cubic interaction can be approximately constructed by repeatedly using linear and OAT interactions. This method should be widely applicable to various spin systems, as the OAT interactions have been experimentally realized in a number of physical systems \cite{riedel2010atom,gross2010nonlinear,hosten2016quantum}. Another one uses light-mediated interactions. The spin system is placed inside an one-side optical cavity, forming a spin-cavity system. We show that, by simply sending an optical pulse, off-resonant with cavity mode, into the spin-cavity system, the cubic NSS dynamics is realized after the reflection of the pulse by the one-sided cavity. Such method should be able to realize the cubic interaction in just one step, which is rather attractive from the perspective of experimental implementation.

Finally, we analyze the impact of spin damping, including the single-spin decay and the collective-spin dephasing. We reveal that,  in the presence of damping, the cubic scheme works much better than the quadratic one. That is, in the weak-coupling regime, the cubic scheme can still maintain is speed advantage in QFI production; besides, the macroscopic superposition state created by the cubic interaction is much more
robust against decoherence than the one created by the quadratic interaction.

The rest of the paper is organized as follows. In \cref{sec:entanglement} we introduce the multipartite entanglement of the collective spins and its correlations with QFI. In \cref{sec:cubicinteraction} we first analytically derive the amount of achievable QFI in the weak coupling regime. Then, we analysis the properties of the macroscopic superposition states created by the cubic interaction. In \cref{sec:evenodd} we discuss the entanglement even-odd effect. In \cref{sec:speedingup} we describe how to speed up the procedure of GHZ-state generation. In \cref{sec:implementation} we present two approaches to realize the cubic NSS dynamics. In \cref{sec:noise} we analysis the impact of the decoherence to the entanglement generation. Finally, we summarize in \cref{sec:conclution}.

%%%%%%%%%%%%%%%%%%%%%%%%%%%%%%%%%%%%%%%%%%%%%%%%%%%%%%%%%%%%%%%%%%%%%%%%%%%%%%%%%%%%%%%%%%%%%%%%%%%%%%%%%%%%%%%%%%%%%%%%%%%%%

\section{Multipartite entanglement in quantum spin systems}
\label{sec:entanglement}
We consider creating multiparticle entanglement among spins in an ensemble consisting of $N$ identical two-level atoms with the excited state $\ket{\uparrow}$ and the ground state $\ket\downarrow$. To describe the collective properties of such system, we define the pseudo angular momentum operators $S_i=\sum_{k} \sigma_{k}^{i}/2(i=x, y, z)$ for atoms, which satisfy the commutation relations  $\left[ {{S_i},{S_j}} \right] = i{\varepsilon _{ijk}}{S_k}$, with ${\varepsilon _{ijk}}$ being the Levi-Civita symbol, where $\sigma_{k}^{i}$ is a Pauli matrix for the $i$th atom, e.g., $\sigma_{x}^{i}=|\uparrow\rangle_{i}\left\langle\left.\downarrow\right|_{i}+\mid \downarrow\right\rangle_{i}\left\langle\left.\uparrow\right|_{i}\right.$. Suppose that all the elementary spins point in the same mean direction $(\theta,\phi)$, that is, each atom is prepared in the state $\left|\theta, \phi\right\rangle_{i}=\cos \frac{\theta}{2}|\uparrow\rangle_{i}+e^{i \phi} \sin \frac{\theta}{2}|\downarrow\rangle_{i}$, forming the well-known coherent spin state (CSS) \cite{PhysRevA.47.5138}
\begin{eqnarray}
\left| {\theta ,\phi } \right\rangle  &=& \left|\theta, \phi\right\rangle_{i}^{\otimes N}=\sum\limits_{k = 0}^{2S} {\sqrt {\frac{{(2S)!}}{{(2S - k)!k!}}} } {}\nonumber\\
&&\times  {\left( {\sin \frac{\theta }{2}} \right)^{2S - k}}{\left( {\cos \frac{\theta }{2}} \right)^k}e^{ik\phi }\left|S, {S - k} \right\rangle\label{eq1},
\end{eqnarray}
where the collective angular momentum states $\ket{S,m}$  (Dicke states) is the eigenstate of $S_z$, satisfying $S_z\ket{S,m}=m\ket{S,m}$ with $S=N/2$. The CSSs are  separable (nonentangled), and a conventional way to entangle the particles is to utilize the second order nonlinear processes, e.g., OAT evolution $U_{\rm{OAT}}=\exp[-i\chi t S_x^2]$ \cite{PhysRevA.47.5138}, where $\chi$ is the coupling constant. To show how spin entanglement is created by $U_{\rm{OAT}}$, assume that the collective spin is polarized along the $z$ direction, leading to the initial state $\ket{\Psi_A}_{\rm{in}}=\ket{\uparrow}^{\otimes N}$. At short times, the evolution of this state is found to be
\begin{eqnarray}
{\left| {{\Psi _A}} \right\rangle _{\rm{out}}}{\rm{ }} &=& {U_{\rm{OAT}}}{\left| {{\Psi _A}} \right\rangle _{\rm{in}}}{\rm{ }}
\nonumber\\&\approx & \mathcal{N}\left({{{\left|  \uparrow  \right\rangle }^{ \otimes N}} - \frac{{2i\alpha }}{{N\left( {1 - i\alpha } \right)}}\sum\limits_{i \ne j} {\left| {{ \downarrow _i}{ \downarrow _j}} \right\rangle } \ket{\uparrow} _{ \ne i,j}^{ \otimes \left( {N - 2} \right)}} \right),\nonumber \end{eqnarray}
where $\mathcal{N}=(1-i\alpha)/\sqrt{1+3\alpha^2}$ is a normalization constant with $\alpha= N\chi t$, and in deriving the last equality we have kept terms up to first order in $S_x^2$ and used the relations $\sigma^x_i\ket{\downarrow}_i(\ket{\uparrow}_i)=\ket{\uparrow}_i(\ket{\downarrow}_i)$. Obviously, the entanglement between the initial and first coupled (double-spin-flipped) states has been created. Such pairwise entanglement have garnered tremendous attention for many years \cite{ma2011quantum}, as they are the origin of spin squeezing, which have important applications in quantum metrology as well as in fundamental physics \cite{PhysRevA.47.5138,PhysRevA.50.67}. In fact, irrespective of the creation of spin squeezing, multiparticle entanglement can also be produced by higher-order nonlinearity, such as the three-order (cubic) evolution $U_{x}=\exp\{-i\chi t S_x^3\}$. For this evolution, one may also derive the time evolved state at time $t$
\begin{eqnarray}
{\left| {{\Psi _A}} \right\rangle _{\rm{out}}}{\rm{ }} &=& {U_x}{\left| {{\Psi _A}} \right\rangle _{\rm{in}}}{\rm{ }} \nonumber\\&\approx& \mathcal{N}\left( {{{\left|  \uparrow  \right\rangle }^{ \otimes N}} - \frac{{3i\alpha }}{4N}\sum\limits_{i \ne j \ne k} {\left| {{ \downarrow _i}{ \downarrow _j}{ \downarrow _k}} \right\rangle } | \uparrow \rangle _{ \ne i,j,k}^{ \otimes \left( {N - 3} \right)}} \right)\nonumber
\end{eqnarray}
with the  normalization constant $\mathcal{N}=1/\sqrt{1+3N\alpha^2/32}$, showing that the triple-wise entanglement among spins is produced. Obviously, such a state exhibits no property of spin squeezing \cite{PhysRevA.68.012101}, while a natural question arises: is it useful for sub-shot-noise interferometry?

To answer this question we use the QFI to quantify the degree of useful entanglement for quantum metrology. The QFI is closely related to the multipartite entanglement \cite{PhysRevA.85.022321} and also gives the fundamental limit to the precision achievable in an unknown-parameter  estimation protocol \cite{PhysRevLett.102.100401}. Considering a scenario of phase estimation, a probe spin state $\rho_{\rm{in}}$ is transformed into ${\rho _\beta } = \exp \left( { - {\rm{i}}\beta {S_{\bm{n}}}} \right){\rho _{{\rm{in }}}}\exp \left( {{\rm{i}}\beta {S_{\bm{n}}}} \right)$ by the $\bm{n}$-direction collective spin generator $S_{\bm{n}}$, where $\beta$ denotes an unknown phase shift to be estimated. The phase sensitivity is limited by the quantum Cram$\acute{e}$r-Rao bound \cite{helstrom1976.123}:
\begin{equation}
	\Delta {\beta} \ge \Delta {\beta_{\rm{QCR}}} =\frac{1} {{\sqrt {{F_Q}\left[ {\rho_{\rm{in}}, S_{\bm{n}}} \right]} }}\label{eq2},
\end{equation}
where
\begin{equation}
	F_{Q}[\rho, {S_{\bm{n}}}]=2 \sum_{l, l^{\prime}} \frac{\left(\lambda_{l}-\lambda_{l^{\prime}}\right)^{2}}{\lambda_{l}+\lambda_{l^{\prime}}}\left|\left\langle l|{J}| l^{\prime}\right\rangle\right|^{2}\label{eq3}
\end{equation}
is the QFI, $\lambda_l$ and $\ket{l}$ are the eigenvalues and eigenvectors of the probe state $\rho_{\rm{in}}$, respectively. The QFI is a measure
of how susceptible of $\rho_{\rm{in}}$ to small influences induced by $S_{\bm{n}}$. The larger the value of QFI, the more precision the estimation. In the case of pure state, $\rho_{\rm{in}}=\ket{\psi_{\rm{in}}}\bra{\psi_{\rm{in}}}$, Eq. (\ref{eq3}) can be further simplified to \cite{PhysRevLett.72.3439}
\begin{equation}
	F_{Q}[\rho_{\rm{in}} , {S_{\bm{n}}}]=4(\Delta {S_{\bm{n}}})^{2}_{\ket{\psi_{\rm{in}}}}\label{eq4},
\end{equation}
where ${(\Delta A)^2_{\ket{\psi}}} = \langle\psi| {{A^2}} |\psi\rangle  - {\langle \psi|A|\psi\rangle ^2}$ is the variance of $A$ in the state $\ket{\psi}$.
For a given probe state $\rho_{\rm{in}}$, it is needed to optimize the rotation direction, $\bm{n}\rightarrow \bm{n}_{\rm{op}}$, to maximize the variances of $S_{\bm{n}}$ (thus QFI) \cite{RevModPhys.90.035005}. If, for example, the probe state is in the separable CSS $\ket{\psi_{\rm{in}}}=\ket{\frac{\pi}{2},0}$, one may choose $\bm{n}_{\rm{op}}=z$ to yield $F_Q=N$, resulting in a sensitivity $\Delta\beta=1/\sqrt{N}$, which is exactly the SQL mentioned above. To overcome this limit, one should use the entangled states, e.g., the GHZ states \cite{GHZ}, $\ket{\psi_{\rm{in}}}=\frac{1}{\sqrt{2}}(\ket{\frac{\pi}{2},0}+\ket{\frac{\pi}{2},\pi})$, with which the QFI can be calculated (by choosing $\bm{n}_{\rm{op}}=x$) to give $F_Q=N^2$, leading to the HL sensitivity $\Delta\beta=1/N$. One thus can conclude that any entangled states whose QFI satify $N<F_Q\leq N^2$ are useful for sub-SQL sensitivity \cite{PhysRevLett.102.100401}.

\section{The cubic interactions}
\label{sec:cubicinteraction}
\subsection{Weak coupling regime}
\begin{figure}[tp]
	\centering
	\includegraphics[scale=0.64]{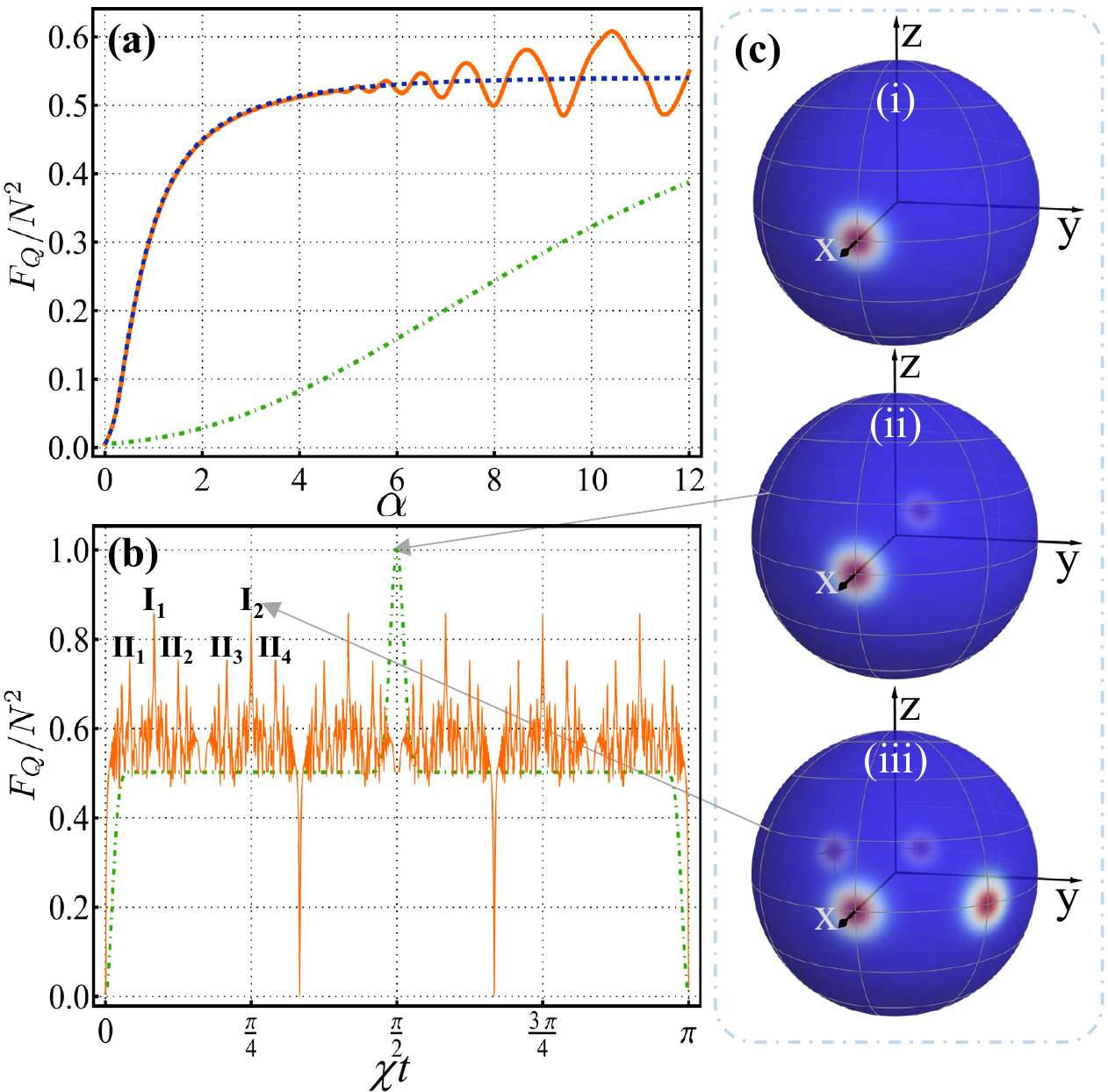}
	\caption{(Color online) QFI of the time-evolved states versus coupling strength (expressed in terms of either $\alpha$ or $\chi t$, see text for clarification) for $N=200$: exact numerical solution of the cubic scheme (solid orange curve) and the OAT scheme (dash-dotted
 green curve). (a) The dashed blue curve is the analytical result of Eq. (\ref{eq16}). (b) The peaks marked $\text{I}_{1,2}$ are the maximum QFI that is achievable by the cubic scheme, while the submaximal QFI are marked by $\text{II}_{1-4}$. (c) Quasiprobability distribution of different spin states: (i) is a CSS, (ii) is a GHZ state, and (iii) is a four-components Schr$\ddot{\text{o}}$dinger cat state. }
\label{fig1}
\end{figure}

We now proceed with the derivation of the QFI of the cubic-interaction-evolved states. For convenience, we suppose that the spins are initially prepared  in the CSS, $\ket{\psi}=\ket{\frac{\pi}{2},0}$, which is subjected to the time evolution
\begin{equation}
	{U_z} = \exp \left( { - i\chi tS_z^3} \right).\label{eq5}
\end{equation}
One thus obtains the probe state at time $t$
\begin{eqnarray}
\left| {{\psi _{\rm{in}}}} (t)\right\rangle  = {U_z}\left|\psi \right\rangle  &=&\frac{1}{2^S} \sum\limits_{k = 0}^{2S} {\sqrt {\frac{{(2S)!}}{{(2S - k)!k!}}} }\nonumber\\
 &&\times{e^{ - i\chi t{{\left( {S - k} \right)}^3}}}\left| {S,S - k} \right\rangle.\label{eq6}
\end{eqnarray}
For this state, since $[S_z,U_z]=0$, $S_z$ is conserved during evolution. Therefore, the uncertainties are redistributed only in the $x$-$y$ plane [see \cref{fig1}(c)], which predicts that the optimal direction of the generator,  $\bm n_{\rm{op}}$, is in some direction in the $x$-$y$ plane. To see how the uncertainties are redistributed, we next work in the Heisenberg picture. The time evolution of the ladder operators ${S_ \pm } = {S_x} \pm i{S_y}$ can be exactly evaluated to give \cite{PhysRevA.47.5138}:
\begin{equation}
{S_ - }(t) = U_z^\dag {S_ - }(0){U_z} = {{\rm{e}}^{ - i\mu \left( {{S_z}^2 + {S_z} + \frac{1}{3}} \right)}}{S_ - }(0),\label{eq7}
\end{equation}
where $\mu\equiv3\chi t$. The transverse components after the cubic evolution are then given by
\begin{eqnarray}
{S_x}(t) = \frac{1}{2}\left[ {{S_ + }{{\rm{e}}^{i\mu \left( {S_z^2 + {S_z} + \frac{1}{3}} \right)}} + {{\rm{e}}^{ - i\mu \left( {S_z^2 + {S_z} + \frac{1}{3}} \right)}}{S_ - }} \right],\label{eq8}\\
{S_y}(t) = \frac{1}{{2i}}\left[ {{S_ + }{{\rm{e}}^{i\mu \left( {S_z^2 + {S_z} + \frac{1}{3}} \right)}} - {{\rm{e}}^{ - i\mu \left( {S_z^2 + {S_z} + \frac{1}{3}} \right)}}{S_ - }} \right].\label{eq9}
\end{eqnarray}
To find $\bm n_{\rm{op}}$, we calculate the variance of an arbitrary angular
momentum operator along the $\phi$ direction, ${S_{\phi}} = {S_x}(t)\cos \phi  + {S_y}(t)\sin \phi $, in the $x$-$y$ plane, yielding
\begin{eqnarray}
{(\Delta {S_\phi })^2_{\ket{\psi}}} &=& {\cos ^2}\phi {\left( {\Delta {S_x}} \right)^2_{\ket{\psi}}} + {\sin ^2}\phi {\left( {\Delta {S_y}} \right)^2_{\ket{\psi}}}\nonumber\\
 &&+ \sin2\phi\left(\frac{1}{2}\left\langle {{{\{ {{S_x},{S_y}} \}} }} \right\rangle-\langle S_x\rangle\langle S_y\rangle\right),\label{eq10}
\end{eqnarray}
where $\{.,.\}$ denotes the anticommutator of two observables. To calculate the first moments of the spin components in Eq. (\ref{eq10}), we turn to evaluate the mean of the ladder operator
\begin{eqnarray}
\left\langle {{S_ + }\left( t \right)} \right\rangle  &=& {2^{ - 2S}}\sum\limits_{k = 0}^{2S} {\sum\limits_{l = 0}^{2S} {\sqrt {\frac{{\left( {2S} \right)!}}{{\left( {2S - k} \right)!k!}}} } } \sqrt {\frac{{\left( {2S} \right)!}}{{\left( {2S - l} \right)!l!}}} \nonumber\\
 &&\times \langle S,S - k|{S_ + }{{\rm{e}}^{i\mu \left( {{S_z^2} + {S_z} + 1/3} \right)}}\left| {S,S - l} \right\rangle \nonumber\\
 &=& {2^{ - 2S}}\sum\limits_{l = 1}^{2S} {\frac{{\left( {2S} \right)!}}{{\left( {2S - l} \right)!(l - 1)!}}} {{\rm{e}}^{i\mu \left[ {{{\left( {S - l} \right)}^2} + S - l + \frac{1}{3}} \right]}}\nonumber\\
 &=& S\sum\limits_{m = 0}^{2S - 1} {P_{2S - 1}}\left( m \right){{\rm{e}}^{i\mu \left[ {{{( {S - m } )}^2} -(S - m) + \frac{1}{3}} \right]}},\label{eq11}
\end{eqnarray}
where in the last equality we set $m=l-1$ and the binomial distribution $P_{2S - 1}( m )$ can be approximately converted to the Gaussian distribution,
\begin{eqnarray}
{P_{2S - 1}}\left( m \right) &=& \frac{{\left( {2S - 1} \right)!}}{{\left( {2S - 1 - m} \right)!m!}}{\left( {\frac{1}{2}} \right)^{2S - 1 - m}}{\left( {\frac{1}{2}} \right)^m}\nonumber\\
 &\simeq & \frac{1}{{\sqrt {S\pi } }}{{\rm{e}}^{ - {{\frac{{\left( {S - m} \right)}}{S}}^2}}},\label{eq12}
\end{eqnarray}
for large $S$ (see Appendix \ref{App:a} for details). We thus obtain
\begin{eqnarray}
\left\langle {{S_ + }\left( t \right)} \right\rangle  &=& \frac{S}{{\sqrt \pi  }}\frac{1}{{\sqrt S }}\sum\limits_{k =  - \sqrt S }^{\sqrt S } {{{\rm{e}}^{ - \left( {1 - i\mu S} \right){k^2} - i\mu \sqrt S k + \frac{1}{3}i\mu }}} \nonumber\\
 &\simeq& \frac{S}{{\sqrt \pi  }}\int_{ - \infty }^{ + \infty } {{{\rm{e}}^{ - \left( {1 - i\mu S} \right){k^2} - i\mu \sqrt S k + \frac{1}{3}i\mu }}dk} \nonumber\\
 &\simeq& \frac{S}{{\sqrt {1 - i\mu S} }},\label{eq13}
\end{eqnarray}
where $k=(S-m)/\sqrt{S}$, and, in the second equality, we have transformed the sum to integral, which is valid only when $\Delta k=1/\sqrt{S}\rightarrow 0$ for, again, large $S$, and, in the last equality, we also have used the approximation $
\exp\mathrm{[}i\mu (4-i\mu S{{)}/{12}}(1-i\mu S)]\approx 1$ for $\mu\ll1$. Along the same lines, one may derive the quadratic expectation values $\langle S_+^2(t)\rangle$ and $\langle S_+(t)S_-(t)\rangle$ (see Appendix \ref{appb} for more details), with which we are able to calculate the means
\begin{eqnarray}
&&\left\langle {{S_x}} \right\rangle  = S\sqrt {{{{\alpha _1}\left( {{\alpha _1} + 1} \right)}}/2} ,\nonumber\\&&\left\langle {{S_y}} \right\rangle  = S\sqrt {{{{\alpha _1}\left( {1-{\alpha _1}} \right)}}{/2}} ,\nonumber\\
&&\left\langle {S_x^2} \right\rangle  = \frac{S}{4}\left[ {\left( {2S + 1} \right) + \left( {2S - 1} \right)\sqrt {{{{\alpha _4}\left( {{\alpha _4} + 1} \right)}}{/2}} } \right],\nonumber\\
&&\left\langle {S_y^2} \right\rangle  = \frac{S}{4}\left[ {\left( {2S + 1} \right) - \left( {2S - 1} \right)\sqrt {{{{\alpha _4}\left( {1 - {\alpha _4}} \right)}}{/2}} } \right],\nonumber\\
&&\left\langle {{{\{ {{S_x},{S_y}} \}}}} \right\rangle  =  \frac{S}{2}\left( {2S -1} \right)\sqrt {{{{\alpha _4}\left( {1 - {\alpha _4}} \right)}}{/2}},\label{eq14}
\end{eqnarray}
where we have defined the new parameters $\alpha_k=1/\sqrt{1+kS^2\mu^2}$ with $k=1,...,4$. Substituting these values into Eq. (\ref{eq10}) we finally arrive at
\begin{eqnarray}
(\Delta S_{\phi})^2_{\ket{\psi}}&=&\frac{S}{4}\left[ 2\left( 1-\alpha _1 \right) S+1 \right]
\nonumber\\
&&+\sqrt{\mathcal{A} ^2+\mathcal{B} ^2}\cos \left( 2\phi -2\delta \right)
,\label{eq15}
\end{eqnarray}
where
\begin{eqnarray}
\mathcal{A}&=&\frac{S}{4\sqrt{2}}(2S-1)\sqrt{\alpha _4\left( 1+\alpha _4 \right)}-\frac{S^2}{2}\alpha _{1}^{2},\nonumber
\\
\mathcal{B}&=&\frac{S}{4\sqrt{2}}\left( 2S-1 \right) \sqrt{\alpha _4\left( 1-\alpha _4 \right)}-\frac{\mu S^3}{2}\alpha _{1}^{2},\nonumber\\
\delta&=&\frac{1}{2}\arctan\left(\frac{\mathcal{B}}{\mathcal{A}}\right).\nonumber
\end{eqnarray}
Eq. (\ref{eq15}) is maximized when $\phi=\delta$, obtaining
\begin{eqnarray}
F_Q=4\sqrt{\mathcal{A}^2+\mathcal{B}^2}+S\left[ 2\left( 1-\alpha _1 \right) S+1 \right].\label{eq16}
\end{eqnarray}
For $S\mu\ll 1$, Eq. (\ref{eq16}) can be approximated as
\begin{eqnarray}
F_Q\approx 2S+\frac{9}{2}S^2\alpha ^2\geq N,
\end{eqnarray}
which indicates that any nonzero $\alpha$ enables the sensitivity to surpass the SQL. Therefore, the entanglement created by Eq. (\ref{eq5}) is \emph{useful} for quantum metrology. For comparison, the QFI created by OAT interaction in the weak coupling regime is also calculated to give: $F_{Q_{\rm{OAT}}}\approx 2S+2S\alpha ^2$. Apparently, the QFI produced by cubic interaction is about $S$ times faster than OAT interaction. This is quite a promising advantage, since the ability to create entangled quantum resources rapidly is a pursuit in quantum metrology. It should be emphasized that the speed-up rate is closely connected to $N$ (the large the $N$, the faster the increase in QFI), which means that the cubic scheme might be more suitable for atomic systems with a large number of atoms \cite{Nature.581.7807,PhysRevLett.102.033601,PhysRevLett.104.073602,PhysRevLett.122.223203,PhysRevLett.122.223203}.

In \cref{fig1}(a) we compare the analytical result of Eq. (\ref{eq16}) (dashed blue curve) and the exact numerical results from Eq. (\ref{eq6}) (solid orange curve). The two curves fit pretty well in the weak coupling regime and gradually deviate when $\alpha$ increases. For large $\alpha$, the numerical results display various oscillating structures, which are lost by analytical result due to the discrete-to-continuous conversion in Eq. (\ref{eq13}). In fact, each peak of QFI is related to a macroscopic supposition of collective spin, as will be discussed below.
  \cref{fig1}(a) also confirms that the QFI of the cubic scheme increases much more rapidly with $\alpha$ than the OAT scheme (dash-dotted green curve). In \cref{fig1}(b) we also plot the periodic evolution of QFI in time for both schemes. It shows that the OAT scheme can saturate the HL at $t=\pi/2\chi$, which corresponds to the creation of a GHZ state \cite{PhysRevA.56.2249}, as shown in \cref{fig1}(c)(ii). The QFI of the cubic scheme, however, has a quite complicated structure. Although can not saturate the HL, there exist two maximum peaks [labeled by $\text{I}_{1,2}$ in \cref{fig1}(b)] that are quite near the HL in a period of evolution, which corresponds to a Schr$\ddot{\text{o}}$dinger cat state with four superposed CSSs [see \cref{fig1}(c)(iii)]. Besides the two maximum peaks, there also exist a number of lower peaks, e.g., four secondary peaks [labeled by $\text{II}_{1-4}$ in \cref{fig1}(b)]. Next, we quantify the amount of QFI for these peaks and explore the properties of these peak states .

 \subsection{Strong coupling regime}
 %\subsubsection{Even-N spin states}
 For convenience, we first assume that $N$ is even and rewrite the state of Eq. (\ref{eq6}) in the following form
\begin{eqnarray}
\left| {{\psi _{\rm{in}}}}(t)\right\rangle  &=& \frac{1}{2^S}\sum\limits_{m =  - S}^S {\sqrt {\frac{{(2S)!}}{{(S + m)!(S - m)!}}} } {e^{ - i\chi t{m^3}}}\left| {S,m} \right\rangle\label{eq18}\nonumber\\
\end{eqnarray}
by setting $k=S-m$. Considering the time evolved state at special time $t=\pi/n\chi$ \cite{1993Production}, where $n$ is an integer, the evolution factor $\exp(-i\pi m^3/n)$ at this time has the following periodic properties:
\begin{eqnarray}
\exp \left[ { - \frac{i\pi }{n}{{\left( {m + 2n} \right)}^3}} \right] &=& \exp \left( { - \frac{i\pi }{n}{m^3}} \right).\label{eq19}
\end{eqnarray}
Such periodicity property enables us to expand the evolution factor as a  Fourier series \cite{1993Production}
\begin{eqnarray}
\exp \left[ { - \frac{{i\pi }}{n}{m^3}} \right] = \sum\limits_{q = 0}^{2n - 1} {f_q^e} \exp \left[ { - \frac{{i\pi q}}{n}m} \right],\label{eq20}
\end{eqnarray}
where the coefficients $f_q^e$ are given by the inverse Fourier transform
\begin{eqnarray}
f_q^{e} = \frac{1}{{2n}}\sum\limits_{m = 0}^{2n - 1} {\exp \left[ {\frac{{i\pi q}}{n}m} \right]\exp \left[ { - \frac{{i\pi }}{n}{m^3}} \right]} .\label{eq21}
\end{eqnarray}
Eq. (\ref{eq20}) indicates that we have successfully converted an exponentially cubic form into sums of exponentials linear in $m$, which is a key step for the derivation.
Inserting Eq. (\ref{eq20}) into Eq. (\ref{eq18}), we obtain
\begin{eqnarray}
\left| {{\psi _{\rm{in}}}\left( {\frac{\pi }{{n\chi }}} \right)} \right\rangle  = \sum\limits_{q = 0}^{2n - 1} {f_q^e} \left| {\frac{\pi }{2},  \frac{{\pi q}}{n}} \right\rangle,\label{eq22}
\end{eqnarray}
which shows that a Schr\"{o}dinger-cat-like state (SCS) (a superposition of the CSSs) can be produced by the cubic evolution at the particular time $t=\pi/n\chi$. The characteristics of the SCSs are determined by the coefficients $f_q^e$. Specifically, by using Eqs. (\ref{eq21}) and (\ref{eq22}), one may derive the form of SCS for $n=4$,
\begin{figure*}[tp]
	\centering
	\includegraphics[scale=0.66]{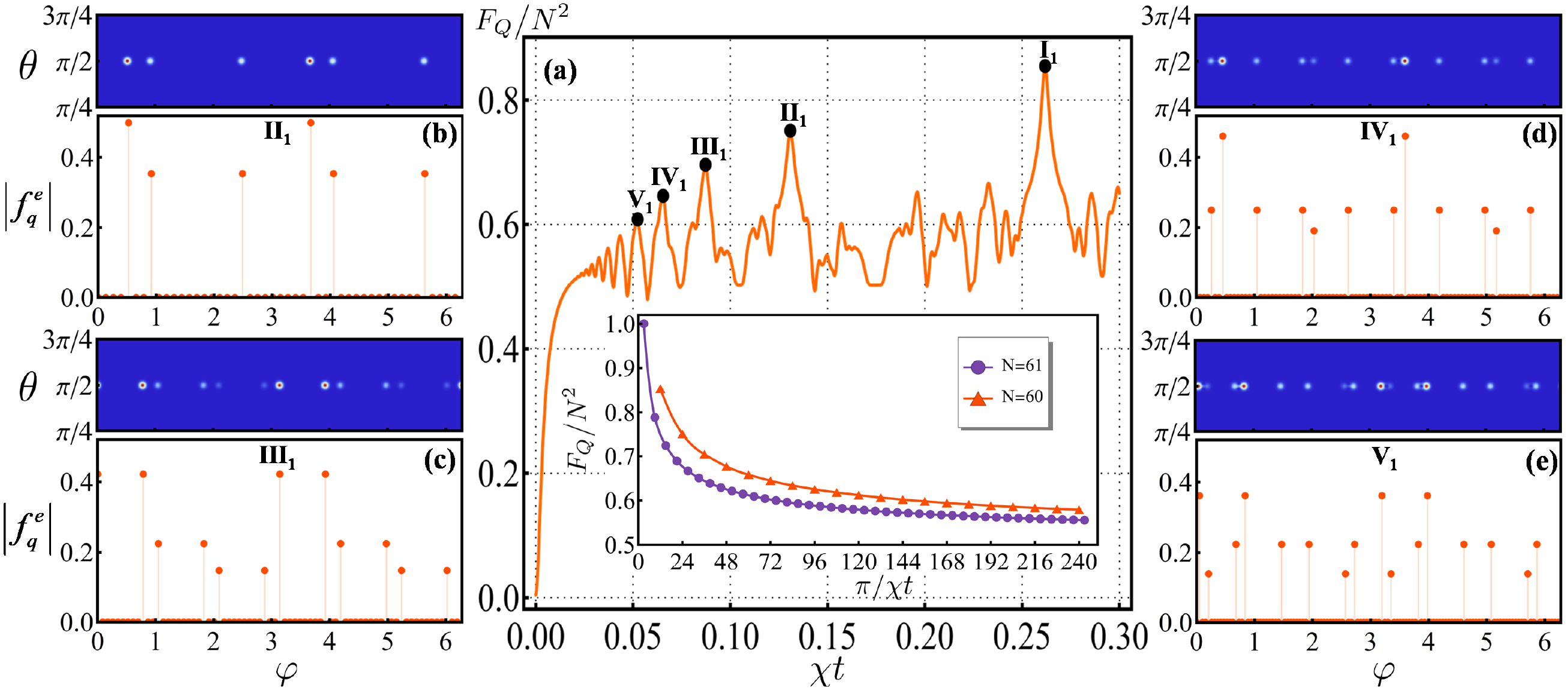}
	\caption{(Color online) (a) QFI produced by the cubic scheme versus coupling strength $\chi t$ for $N=200$. The QFI peaks marked $\text{I}_1,...,\text{V}_1$ are produced by the states $\ket{\psi_{\rm{in}}({\pi}/{12k\chi})}$ in Eq. (\ref{eq6}) with $k= 1,...,5$, respectively. Insert: The QFI of the states $\ket{\psi_{\rm{in}}({\pi}/{12k\chi})}$ and $\ket{\psi_{\rm{in}}({\pi}/{3(2k-1)\chi})}$ vs $k$ for $N=60$ and $N=61$, respectively. (b)--(e) The Fourier-coefficients distribution [ given by Eq. (\ref{eq22})] (bottom) and the QDP (top) of the states of the peaks marked $\text{II}_1-\text{V}_1$ in (a) for $N=1500$.  }
\label{fig2}
\end{figure*}
\begin{table*}
\caption{\label{tab:table1}%
		The quantum state, QFI [calculated via Eq. (\ref{eq33})], and achievable sensitivity of each peak in \cref{fig2}(a).
	}
     \begin{tabular}{p{1cm} p{12cm} p{3cm} p{1.2cm} p{0.0000001cm}}

       \hline \hline

 \centering Peaks&\centering The quantum state of peaks (neglecting the normalization) &\centering QFI&\centering$\Delta\beta$&\\

    \hline

  \centering $\text{I}_1$ &\centering\begin{scriptsize}$\ket{\psi_{\rm{in}}\left(\frac{\pi}{12\chi}\right)}=\ket{ {{\rm{GHZ}}_{\pi/12}^ - } }  + \ket{ {{\rm{GHZ}}_{\pi /3}^ + } }$\end{scriptsize}&\centering 0.85$N^2$ &\centering$1.08/N$&\\

    \hline

   \centering $\text{II}_1$& \centering\begin{scriptsize}$\begin{array} {r@{}l@{}} \ket{\psi_{\rm{in}}\left(\frac{\pi}{24\chi}\right)}=\ket{ {{\rm{GHZ}}_{\pi /6}^ + } }  + \frac{7}{10}\ket{ {{\rm{GHZ}}_{7\pi /24}^ - } } -\frac{7}{10}\ket{ {{\rm{GHZ}}_{19\pi /24}^ - } } \end{array}$\end{scriptsize}& \centering 0.75$N^2$  &\centering $1.15/N$ &\\
    \hline

   \centering $\text{III}_1$&\centering \begin{scriptsize}$\begin{array} {r@{}l@{}}\ket{\psi_{\rm{in}}\left(\frac{\pi}{36\chi}\right)}=\ket{ {{\rm{GHZ}}_0^ + } }  + \ket{ {{\rm{GHZ}}_{\pi /4}^ - } }  + \frac{1}{2}\ket{ {{\rm{GHZ}}_{\pi /3}^ + } } - \frac{1}{2}\ket{ {{\rm{GHZ}}_{7\pi /12}^ + } }  - \frac{1}{3}\ket{ {{\rm{GHZ}}_{2\pi /3}^ - } }  - \frac{1}{3}\ket{ {{\rm{GHZ}}_{11\pi /12}^ - }}  \end{array}$\end{scriptsize}&\centering 0.70$N^2$  &\centering $1.20/N$ &\\
    \hline

   \centering $\text{IV}_1$&\centering\begin{scriptsize}$\begin{array} {r@{}l@{}}\ket{\psi_{\rm{in}}\left(\frac{\pi}{48\chi}\right)}=\ket{ {{\rm{GHZ}}_{\pi /12}^ + } }  + \frac{9}{5}\ket{ {{\rm{GHZ}}_{7\pi /48}^ - } } + \ket{ {{\rm{GHZ}}_{\pi /3}^ + } }  + \ket{ {{\rm{GHZ}}_{7\pi /12}^ + } } -\frac{4}{5} \ket{ {{\rm{GHZ}}_{31\pi /48}^ - } }  - \ket{ {{\rm{GHZ}}_{11\pi /6}^ - } }\end{array}$\end{scriptsize}&  \centering 0.67$N^2$& \centering $1.22/N$&\\
    \hline

   \centering $\text{V}_1$&\centering\begin{scriptsize}$\begin{array} {r@{}l@{}}\ket{\psi_{\rm{in}}\left(\frac{\pi}{60\chi}\right)}=&\frac{8}{5}\ket{ {{\rm{GHZ}}_{\pi /60}^ - } }  + \frac{3}{5}\ket{ {{\rm{GHZ}}_{\pi /15}^ + } }  + \ket{ {{\rm{GHZ}}_{13\pi /60}^ - } }  + \frac{8}{5}\ket{ {{\rm{GHZ}}_{4\pi /15}^ + } } - \ket{ {{\rm{GHZ}}_{7\pi /15}^ + } }  \\&- \ket{ {{\rm{GHZ}}_{37\pi /60}^ - } }  + \frac{3}{5}\ket{ {{\rm{GHZ}}_{49\pi /60}^ - } }  + \ket{ {{\rm{GHZ}}_{13\pi /15}^ + } }  \end{array}$\end{scriptsize}&\centering 0.65$N^2$& \centering $1.24/N$&\\

    \hline  \hline
  \end{tabular}
    \label{tab:ghz}
\end{table*}

\begin{eqnarray}
\left| {{\psi _{\rm{in}}}\left( {\frac{\pi }{{4\chi }}} \right)} \right\rangle &=&\frac{1}{2}\left( {\left| {\frac{\pi }{2},0} \right\rangle  + \left| {\frac{\pi }{2},\frac{\pi }{4}} \right\rangle } \right.\nonumber\\
&&\left. { + \left| {\frac{\pi }{2},\pi } \right\rangle  - \left| {\frac{\pi }{2},\frac{{5\pi }}{4}} \right\rangle } \right),\label{eq23}
\end{eqnarray}
and for $n=12$,
\begin{eqnarray}
\left| {{\psi _{\rm{in}}}\left( {\frac{\pi }{{12\chi }}} \right)} \right\rangle &=&\frac{1}{2}\left( {\left| {\frac{\pi }{2},\frac{\pi }{12}} \right\rangle  + \left| {\frac{\pi }{2},\frac{{\pi }}{3}} \right\rangle } \right.\nonumber\\
&&\left. { - \left| {\frac{\pi }{2},\frac{{13\pi }}{12}} \right\rangle  + \left| {\frac{\pi }{2},\frac{{4\pi }}{3}} \right\rangle } \right).\label{eq24}
\end{eqnarray}
Notably, the states (\ref{eq23}) and (\ref{eq24}) are just the two states that create the two maximum QFI peaks $\text{I}_2$ and $\text{I}_1$ [see \cref{fig1}(b)], respectively. Next, we turn to derive the amount of QFI of peak $\text{I}_2$. By using the state in Eq. (\ref{eq23}), one may directly calculate the means and variances of the collective spin components, obtaining
\begin{eqnarray}
&&\left\langle {{S_x}} \right\rangle  = S{\left( {\cos \frac{\pi }{8}} \right)^{2S - 1}}\cos \left[ {\frac{\pi }{8}\left( {2S + 1} \right)} \right],\nonumber\\
&&\left\langle {{S_y}} \right\rangle  = S{\left( {\cos \frac{\pi }{8}} \right)^{2S - 1}}\sin \left[ {\frac{\pi }{8}\left( {2S + 1} \right)} \right],\nonumber\\
&&\left\langle {S_x^2} \right\rangle  = \frac{1}{8}\left( {6{S^2} + S} \right),\left\langle {S_y^2} \right\rangle  = \frac{1}{8}\left( {2{S^2} + 3S} \right),\nonumber\\
&&\left\langle {{{\{ {{S_x},{S_y}} \}} }} \right\rangle  = \frac{1}{4}\left( {2{S^2} - S} \right).
\end{eqnarray}
Substituting them into Eq. (\ref{eq10}), after optimization of $(\Delta S_\phi)^2_{\ket{ {{\psi _{in}}( {{\pi }/{{4\chi }}} )}}}$ we get
\begin{eqnarray}
{F_Q} &=& 2{S^2}\left\{ {1 + \frac{1}{{\sqrt 2 }} - {{\left( {\cos \frac{\pi }{8}} \right)}^{4S - 2}}} \right.\nonumber\\
&&\left. { \times \left[ {1 + \cos \left( {\frac{{\pi S}}{2}} \right)} \right]} \right\} + \left( {1 - \frac{1}{{\sqrt 2 }}} \right)S\label{eq26}
\end{eqnarray}
for $\phi=\pi/8$. For large $N$, Eq. (\ref{eq26}) is reduced down to
\begin{eqnarray}
F_Q \approx \frac{1}{2}\left( 1+ \frac{1}{{\sqrt 2}}\right)N^2.\label{eq27}
\end{eqnarray}
Eq. (\ref{eq27}) is the upper bound of QFI produced by the cubic scheme in the case of even $N$. Inserting Eq. (\ref{eq27}) into Eq. (\ref{eq2}) yields the best angular
sensitivity achievable by the cubic scheme, $\Delta\beta\simeq1.08/N$, which is very near the HL.

In fact, the Heisenberg scaling of Eq. (\ref{eq27}) originates from the fact that the states of peaks $\text{I}_{1,2}$ are in supposition of two GHZ states, i.e.,
\begin{eqnarray}
\left| {\psi_{\rm{in}}\left(\frac{\pi}{12\chi}\right)} \right\rangle =\frac{1}{\sqrt{2}}\left(\ket{ {{\rm{GHZ}}_{\pi/12}^ - } }  + \ket{ {{\rm{GHZ}}_{\pi /3}^ + } }\right),\label{eq28}
\end{eqnarray}
where we have defined
\begin{eqnarray}
\left| {{\rm{GHZ}}_\varphi ^ \pm } \right\rangle  = \frac{1}{{\sqrt 2 }}\left( {\left| {\frac{\pi }{2},\varphi } \right\rangle  \pm \left| {\frac{\pi }{2},\varphi  + \pi } \right\rangle } \right).\label{eq29}
\end{eqnarray}
Obviously, the maximum-variance direction of the GHZ states of Eq. (\ref{eq29}) is $\varphi$, which from now on we call the direction of a GHZ state. Corresponding to the states (\ref{eq29}), one may derive the variance of $S_\phi$ according to Eq. (\ref{eq10}), yielding
\begin{eqnarray}
(\Delta {S_\phi })_{\left| {{\rm{GHZ}}_\varphi ^ \pm } \right\rangle }^2 &=& \frac{1}{4}\left[ {2{S^2} + S} \right.\nonumber\\
&&\left. { + (2{S^2} - S)\cos 2\left( {\varphi  - \phi } \right)} \right].\label{eq30}
\end{eqnarray}
This equation quantifies the amount of noise in the direction $\phi$ which deviates from the GHZ direction by an angle $\varphi-\phi$. It can maximized to $(\Delta S_{\phi})^2_{\ket{{\rm{GHZ}}_\varphi ^ \pm }}=S^2$ when the two directions are exactly the same ($\phi=\varphi$) and can be minimized to $(\Delta S_{\phi})^2_{\ket{{\rm{GHZ}}_\varphi ^ \pm }}=S/2$ when the two directions are orthogonal to each other ($\phi=\varphi-\pi/2$). For any $0\leq\varphi-\phi\leq\pi/2$, we have $S/2\leq(\Delta S_{\phi})^2_{\ket{{\rm{GHZ}}_\varphi ^ \pm }}\leq S^2$. Therefore, Eq. (\ref{eq30}) could also be regarded as the projection of the noise of a GHZ state in the GHZ direction onto the $\phi$ direction.

Keeping this physical picture in mind, let us turn to seek the maximum variance of the state (\ref{eq28}). By projecting the noise of the two GHZ states, $\ket{ {{\rm{GHZ}}_{\pi/12}^ - } } $ and $\ket{ {{\rm{GHZ}}_{\pi /3}^ + } }$, onto the $\phi$ direction, we get
\begin{eqnarray}
\left(\Delta {S_\phi }\right)_{\left| {{\psi _{\rm{in}}}\left( {\frac{\pi }{{12\chi}}} \right)} \right\rangle }^2 &\approx& \frac{1}{2}\left( {\Delta {S_\phi }} \right)_{\left| {{\rm{GHZ}}_{\pi /12}^ - } \right\rangle }^2 + \frac{1}{2}\left( {\Delta {S_\phi }} \right)_{\left| {{\rm{GHZ}}_{\pi /3}^ + } \right\rangle }^2\nonumber\\
 &=& \frac{1}{4}\left\{ {2{S^2} + S + \frac{1}{2}\left( {2{S^2} - S} \right)} \right.\nonumber\\
&&\left. { \times \left[ {\cos 2\left( {\phi  - \frac{\pi }{{12}}} \right) + \cos 2\left( {\phi  - \frac{\pi }{3}} \right)} \right]} \right\}\nonumber\\
 &=& \frac{1}{4}\left[ {2{S^2} + S + \frac{1}{{\sqrt 2 }}\left( {2{S^2} - S} \right)}\right.\nonumber\\&&\left.\times{\cos \left( {2\phi  - \frac{{5\pi }}{{12}}} \right)} \right],\label{eq31}
\end{eqnarray}
where ${S_{\phi}} = {S_x}(0)\cos \phi  + {S_y}(0)\sin \phi $ and, in the first equality, we have neglected the nondiagonal terms, which is reasonable when $N$ is large (see Appendix \ref{appc} for more details). Eq. (\ref{eq31}) is maximized at $\phi=5\pi/24$ to give $(\Delta {S_{5\pi /24}})_{\left| {{\psi _{\rm{in}}}\left( {\pi /12\chi} \right)} \right\rangle }^2 \approx \frac{1}{2}\left( {1 + \frac{1}{{\sqrt 2 }}} \right){S^2}$, which is exactly the same as the maximum value of  $(\Delta S_\phi)^2_{\ket{ {{\psi _{\rm{in}}}( {{\pi }/{{4\chi }}} )}}}$. Such result (that is, the QFI of $\text{I}_{1,2}$ are equal) has also been predicted by the numerical results in the previous section. This confirms that the noise-projection method outlined above provides a convenient way to derive the QFI of a probe state in a superposition of arbitrary GHZ states, that is,
\begin{eqnarray}
\left| {{\psi _{\rm{in}}}} \right\rangle  = \mathcal{N}\sum\limits_{\varphi } {C_{\varphi } \left| {\rm{GHZ}_{\varphi }^ \pm } \right\rangle },\label{eq32}
\end{eqnarray}
where $\mathcal{N}$ is the normalization and $C_{\varphi}$ are the probability amplitudes. Our task is to maximize the projected noises,
\begin{eqnarray}
(\Delta {S_\phi })_{\left| {{\psi _{\rm{in}}}} \right\rangle }^2 &\approx& {\sum\limits_{{\varphi  }} {\left| {\mathcal{N}{C_{{\varphi  }}}} \right|} ^2}\left( {\Delta {S_\phi }} \right)_{\left| {\rm{GHZ}_{{\varphi }}^ \pm } \right\rangle }^2\nonumber\\
 &=& \frac{1}{4}{\sum\limits_{{\varphi  }} {\left| {\mathcal{N}{C_{{\varphi }}}} \right|} ^2}\left[ {2{S^2} + S} \right.\nonumber\\
&&\left. { + (2{S^2} - S)\cos 2\left( {\varphi  - {\phi  }} \right)} \right],\label{eq33}
\end{eqnarray}
over all values of $\phi$.
 \begin{figure}[tp]
	\centering
	\includegraphics[scale=0.75]{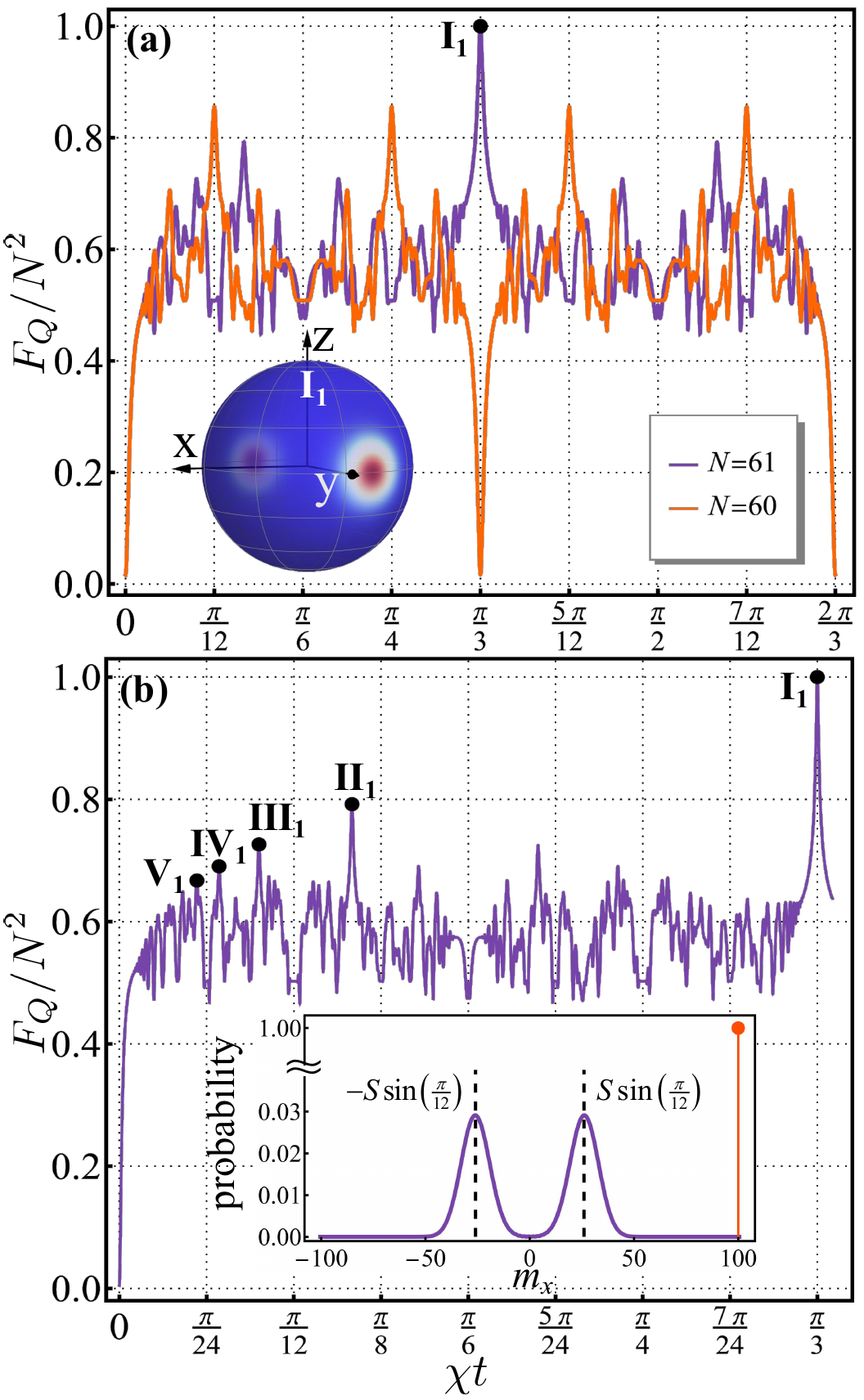}
	\caption{(Color online) (a) QFI produced by the cubic scheme versus coupling strength $\chi t$ for even (orange curve) and odd (purple curve) number of spins. Insert: the QPD of the state corresponds to the peak $\text{I}_1$. (b) QFI produced by the cubic scheme versus coupling strength $\chi t$. The QFI peaks labeled $\text{I}_1,...,\text{V}_1$ are produced by $\ket{\psi_{\rm{in}}({\pi}/{3(2k-1)\chi})}$ with $k= 1,...,5$, respectively. Insert: the $S_x$ probability distributions of the peak GHZ state $\text{I}_1$ (purple curve) and the initial CSS state $\ket{\pi/2,0}$ (orange circle). We here take $N=201$.}
\label{fig3}
\end{figure}

We are now equipped to evaluate the QFI of the lower peaks $\text{II}_{1}-\text{V}_{1}$ [as shown in \cref{fig2}(a)].
In Figs. 2(b)-2(e) we plot the Fourier coefficients distribution as well as the quasiprobability distribution (QPD) [obtained from the exact
numerical evolution given by Eq. (\ref{eq18})] of each peak state, showing that (i)  the two
results are consistent with each other, and (ii) the appearance of large QFI has always been accompanied with the generation of macroscopic quantum-superposition state.  Representing the peak states in terms of the GHZ states of Eq. (\ref{eq29}), we are able to show in table I the explicit forms for each peak state, indicating that they have exactly the same form as the states given in Eq. (\ref{eq32}). One thus can use Eq. (\ref{eq33}) to approximately derive the amount of QFI for each peak for the case of large $N$. As can be seen from table I, the state of each peak could realize a sensitivity near the HL. Interestingly, these peak states appear regularly at a particular time $t_k={\pi}/{12k\chi}$ with integer $k= 1,...,5$. In fact, the states of those not labeled peaks on the left-hand side of the peak $\text{V}_1$ [see \cref{fig2}(a)] are also obtained at $t_k$ but with $k\geq 6$.

It should be emphasized that the number of visible peaks depends heavily on $N$. The larger the $N$, the more QFI peaks it can be seen. This is because the number of CSS components can be found in $\ket{\psi_{\rm{in}}({\pi}/{12k\chi})}$ increases with $k$. For large $k$ but small $N$, these superposition CSS components are overlapped with each other and become indistinguishable. The distance between the CSS components, however, can be increased by increasing $N$, as their distance is proportional to $N$ while the radius of a CSS on the Bloch sphere is proportional to $\sqrt{N}$. Once all the CSS components of $\ket{\psi_{\rm{in}}({\pi}/{12k\chi})}$ become distinguishable on the Bloch sphere, the $k$th QFI peak appears. In other words, the appearance of QFI peaks herald the creation of macroscopic superposition states.
As can be seen from the insert in \cref{fig2}(a), QFI larger than $0.55N^2$ is still obtainable at $t_{240}$ for $N=60$. We emphasize that, although the maximal QFI produced by the cubic scheme is smaller than the maximal QFI of the OAT scheme (that is, $F_Q=N^2$ produced by the GHZ state at time $\pi/2\chi$), it has an advantage of short-preparation time, which, as we will show below, makes the entanglement generation robust against damping.
 \section{the entanglement even-Odd effect}
 \label{sec:evenodd}
An interesting feature of the cubic evolution
of Eq. (\ref{eq5}) is an extreme sensitivity to the parity
of the total spin number $N$. As shown in \cref{fig3}(a), the evolutions of QFI for $N$ atoms vs $N+1$ atoms are
macroscopically different: first, the maximum QFI of the cubic scheme with odd-$N$ spins can saturate to the HL [the peak $\text{I}_1$ in Fig. 3(a)], which is actually produced by the GHZ state $\ket{\rm{GHZ}_{7\pi/12}^-}$ at time $t=7\pi/12\chi$ [see the inset of \cref{fig3}(b)]; second, the state of each QFI peak [the peaks $\text{I}_1-\text{V}_1$  in \cref{fig3}(b)] for odd $N$ appears at a different but regular time $t_k={\pi}/{3(2k-1)\chi}$. These differences originate from the fact that the evolution factor of the odd-N spin system has a quite different periodic property,
\begin{eqnarray}
\exp \left[ { - \frac{i\pi }{n}{{\left( {m + 8n} \right)}^3}} \right] &=& \exp \left( { - \frac{i\pi }{n}{m^3}} \right).\label{eq34}
\end{eqnarray}
Accordingly, the evolution factor can be reexpanded as
\begin{eqnarray}
\exp \left[ { - \frac{{i\pi }}{n}{m^3}} \right] = \sum\limits_{q = 0}^{8n - 1} {f_q^o} \exp \left[ { - \frac{{i\pi q}}{n}m} \right],\label{eq35}\\
f_q^{o} = \frac{1}{{8n}}\sum\limits_{m = 0}^{8n - 1} {\exp \left[ {\frac{{i\pi q}}{n}m} \right]\exp \left[ { - \frac{{i\pi }}{n}{m^3}} \right]}. \label{eq36}
\end{eqnarray}
Then, by using Eqs. (\ref{eq33}) and (\ref{eq36}) we are able to show in the inset of \cref{fig2}(a) (the curve with circle) the amount of QFI for each peak in case of large $N$. Despite the HL QFI peak produced by the state $\ket{\rm{GHZ}_{7\pi/12}^-}$, the spin system with even $N$ is superior over
the one with odd $N$ in the production of multipartite entanglement.
\begin{figure}[bt]
	\centering
	\includegraphics[scale=0.56]{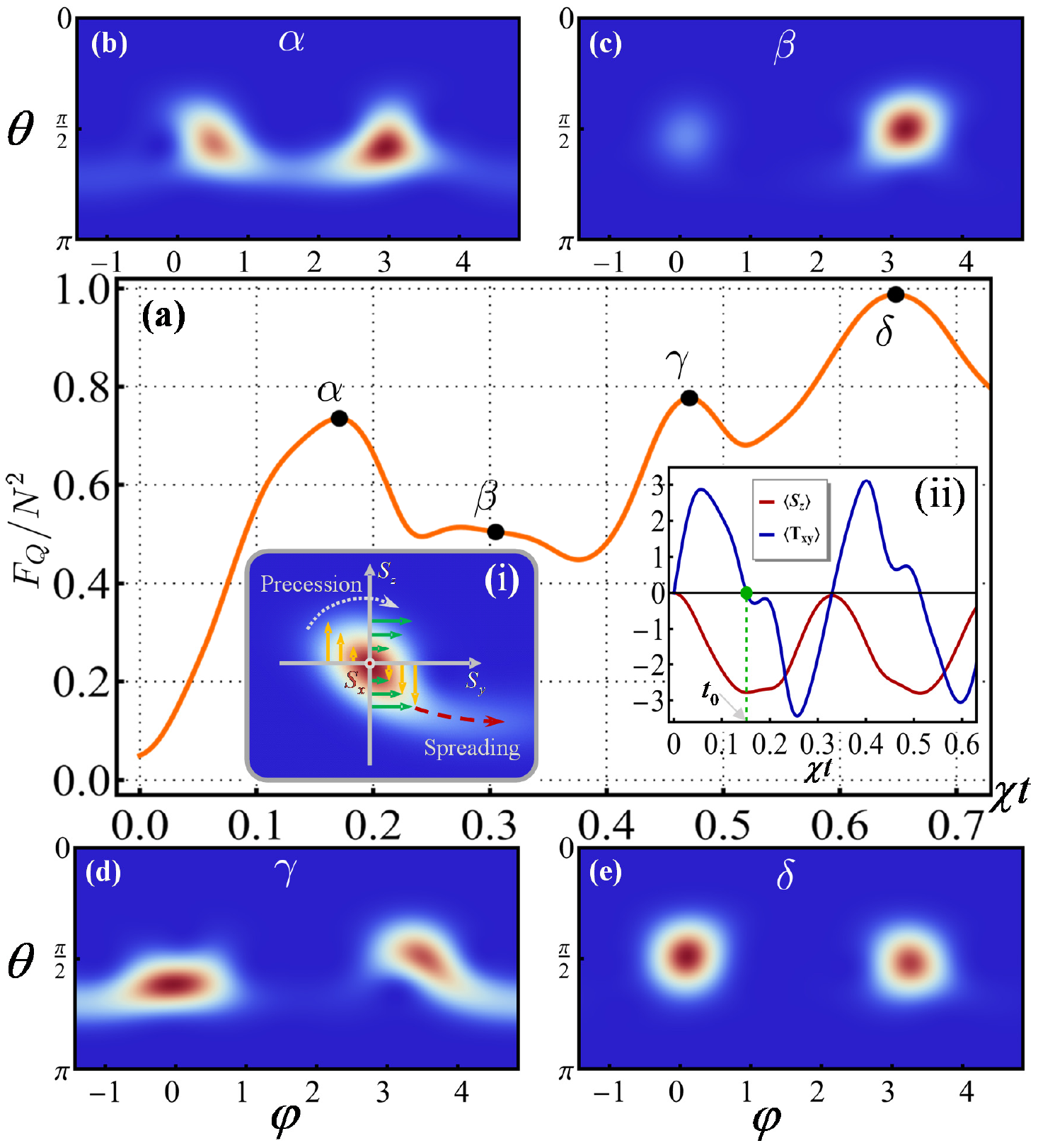}
	\caption{(Color online) (a) QFI produced by the interaction of Eq. (\ref{eq37}) versus coupling strength $\chi t$ for $N=20$. Insert (i): schematic depiction of the evolution induced by the hybrid dynamics on the Bloch sphere, with green flow lines denoting $S_z^2$-dependent rotation of the QPD around the $z$ axis caused by the cubic part in Eq. (\ref{eq37}) and orange flow lines representing twisting of the QPD induced by the quadratic part in Eq. (\ref{eq37}). Insert (ii): we plot the means $\langle T_{{xy}}\rangle$ and $\langle S_{z}\rangle$ as a function of $\chi t$. (b)-(e) are the QPD of the peak states marked with $\alpha,...,\delta$ in (a).     }
\label{fig4}
\end{figure}

A more striking feature is that the amount of QFI at time $t_1=\pi/3$ changes dramatically with the parity of $N$: for even $N$, the evolved spin state is a separable CSS with $F_Q=N$, while, for odd $N$, the spin state created is a maximally entangled GHZ state with $F_Q=N^2$. This entanglement even-odd effect is quite different from the even-odd effect exhibited by the OAT evolution $U_{\rm{OAT}}=\exp[-i\chi t S_z^2]$, which, at the instant $t=\pi/2\chi$, maps the initial CSS $\ket{\pi/2,0}$ to the $N$-dependent GHZ state $\ket{\psi_{\rm{in}}(\pi/2\chi)}=\frac{1}{{\sqrt 2 }}({e^{i\pi /4}}\left| {\pi /2, - \pi (N - 1)/2} \right\rangle  + {e^{ - i\pi /4}}\left| {\pi /2, - \pi (N - 3)/2} \right\rangle )$ \cite{PhysRevA.56.2249,PhysRevLett.82.1835}, showing that the orientation of the created GHZ state is sensitivity to the parity of total spin number $N$.

The determination of the spin number in a realistic quantum system is a critical first step toward the realization of quantum metrology as well as quantum information processing \cite{PhysRevLett.93.143601,PhysRevLett.97.023002,PhysRevLett.98.233601,PhysRevLett.109.133603,PhysRevA.89.043837}. Especially in the context of spin-spin entanglement generation \cite{Nature.438.6453,Science.365.6453}, the spin system will evolve into a highly entangled pure state or a separable mixed state depending on the parity of $N$ \cite{Optics.69,PhysRevA.41.7,PhysRevX.12.011015}. It is thus of great importance to have the ability to determine the parity of $N$ before performing the protocols. Our entanglement even-odd effect might potentially be used
to detect the parity of the total spin number $N$ with a resolution at the single-spin level.
The parity detection proceeds as follows: the spin state is initially prepared in the CSS $\ket{\pi/2,0}$, and then is subjected to the evolution described by Eq. (\ref{eq5}) for a time duration $t=\pi/3\chi$. Finally, measuring $S_x$ a particular outcome $\tilde{S}_x$ is obtained. As can be seen from the inset of Fig. 3(b), for even $N$, we have $\tilde{S}_x=S$, while, for odd $N$, there will be a large probability of finding $\tilde{S}_x$ around $\pm S\sin(\pi/12)$ (especially for the case of large $N$).  Therefore, the measurement of
the angular momentum operator $S_x$ provides a direct way to estimate the parity of $N$.
\begin{figure*}[tp]
	\centering
	\includegraphics[scale=0.65]{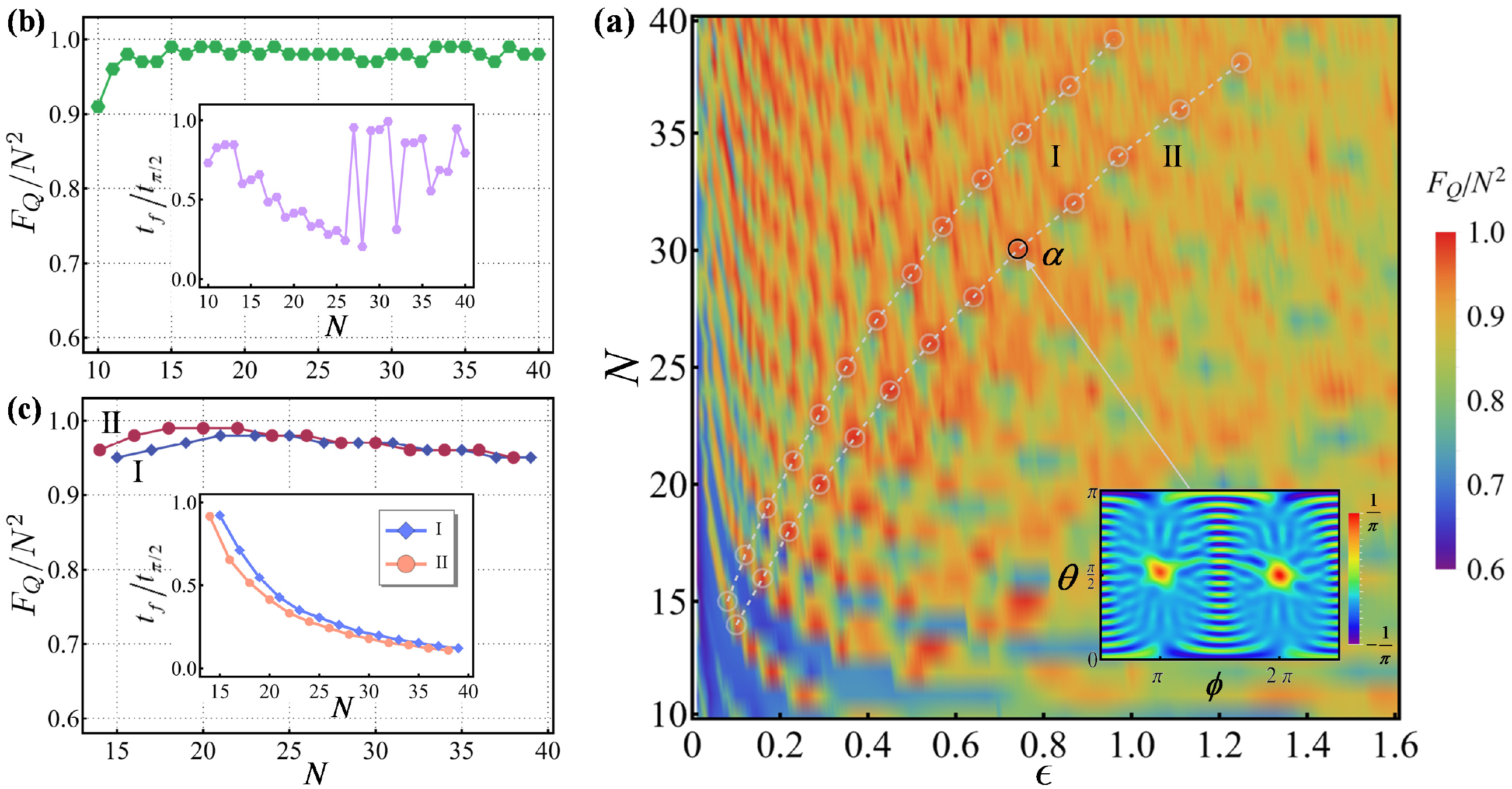}
	\caption{(Color online) (a) Maximally achievable QFI as a function of spin number $N$ and weight factor $\epsilon$. Insert: Wigner quasiprobability distribution $W(\theta,\phi)/\sqrt{2\pi S}$ \cite{PhysRevA.49.4101} of the state at the point marked with $\alpha$. (b) QFI optimized with respect to the coupling strength $\chi t$ and the weight factor $\epsilon$ vs the spin number $N$. Insert: the ratio of the GHZ-state-preparation time $t_f$ of the hybrid scheme to the GHZ-state-preparation time $t_{\frac{\pi}{2}}$ of the OAT scheme as a function of $N$. The procedure of GHZ-state preparation is accelerated if the preparation-time ratio $t_f/t_{\frac{\pi}{2}}<1$. (c) Maximally achievable QFI of the two curves marked I and II in (a) versus $N$. Insert: the preparation-time ratio of the two cures vs $N$.                                                                    } \label{fig5}
\end{figure*}

\section{Speeding up the generation of GHZ states}
\label{sec:speedingup}
The pursuit of the rapid generation of GHZ states is not only vital for quantum metrology but also of fundamental interest in many applications, such as quantum computation \cite{Book1,2021Linear,Barrett_2010,PhysRevA.83.062339}. As was shown previously, the GHZ states can be directly created by applying either the OAT evolution to an even- or odd-$N$ spin system for a time $t_{\frac{\pi}{2}}=\pi/2\chi$ or the cubic evolution to an odd-$N$ spin system for a time $t_{\frac{\pi}{3}}=\pi/3\chi$. However,  in reality, since the coupling constant $\chi$ in most quantum system is normally weak, long preparation time is then required,  which poses a great challenge to the experimental realization (that is, extreme long decoherence time is required when preparing the GHZ state). It is, therefore, highly desirable to develop new protocols that enable the \emph{rapid} creation of GHZ states. We next show how to speed up the procedure of  GHZ-state generation by utilizing a high-order nonlinear spin-spin dynamics.

 Consider a hybrid Hamiltonian of the form
\begin{eqnarray}
H = \chi\left(\epsilon S_z^3{\rm{ + }}S_{{y}}^2\right),\label{eq37}
\end{eqnarray}
which is a combination of the cubic interaction and the OAT interaction with a weight factor $\epsilon$.
If the initial state is prepared in $\ket{\pi/2,0}$, the cubic term $\propto S_z^3$ of Eq. (\ref{eq37}) causes a precession
of the collective spin about the $z$ axis at a rate proportional to $S^2_z$, resulting in a spreading of the QPD along the $S_y$ direction, as shown in the insert (i) of \cref{fig4}(a); analogously, the OAT term $\propto S_y^2$ induces a precession about the $y$ axis at a rate proportional to $S_y$, leading to a spreading of the QPD along either the $+S_z$ or the $-S_z$ direction, depending on the sign of $S_y$. The net effect is that the QPD in the regime $\theta<\pi/2$ precesses around the $x$ axis, analogous to the dynamics induced by the quadratic interaction $S_y^2+S_z^2\propto S_x^2$, and in the regime $\theta>\pi/2$ the QPD is stretched along a direction determined by $\epsilon$, in close analogy with the two-axis twisting dynamics described by $ S_y^2-S_z^2$ \cite{PhysRevA.47.5138}. As a result, the QPD continuously spreads out and gradually arrives at the position $(\pi/2,\pi)$ on the Bloch sphere, forming a Schr\"{o}dinger-cat-like state $\alpha$ that has a quite large amount of QFI, as shown in \cref{fig4}(a) and (b). This state, however, is an imperfect GHZ state as the two superposition components are still overlap with each other. For this state, it is not difficult to find that its expectation $\langle S_z\rangle$ is negative, which evolves according to
\begin{eqnarray}
d\left\langle {{S_z}} \right\rangle /dt =  - \chi \left\langle {{T_{{xy}}}} \right\rangle\label{eq38},
\end{eqnarray}
indicating that the expectation $\langle S_z\rangle$ is determined by the expectation value of the tensor operator ${{T_{xy}} = {S_y}{S_x} + {S_x}{S_y}}$.
 Once the QPD surrounds the Bloch sphere, the sign of $\langle T_{xy}\rangle$ is reversed at time $t=t_0$, as can be seen from the insert (ii) of Fig. 4(a). Then, $\langle S_z\rangle$ starts to increase, which gradually eliminates the QPD between the two superposition components, creating an unequally weighted GHZ state $\beta$ [see \cref{fig4}(c)].  Further evolution of the dynamics of Eq. (\ref{eq37}) will evenly distribute the QPD to the antipodal CSSs [see Fig. 4(d)] and finally produce a near-perfect GHZ state $\delta$ [see \cref{fig4}(e)]. It should be stressed that, given an atom number $N$, the maximum achievable QFI of the GHZ-like state $\delta$ depends heavily on the relative coupling strength $\epsilon$. Taking $N=20$ as an example, the maximum QFI $F_Q/N^2=0.99$ is obtained when $\epsilon=0.29$ at the fixed time $t_f=0.65/\chi$. Obviously, in contrast to the quadratic OAT interaction, the CQA type interaction of Eq. (\ref{eq37}) can speed up  the procedure of GHZ-state generation since the preparation time satisfies $t_f<t_\frac{\pi}{2}$.

In \cref{fig5}(a), we also plot the maximally achievable QFI for different $N$ as a function of the relative coupling strength $\epsilon$. We find that large QFI are  (i) mainly concentrated within the regime of small $\epsilon$ and (ii) more easily achieved when the atom number $N$ is large. \cref{fig5}(b) shows the maximal achievable QFI of the proposed protocol for different choices of $N$. QFI as large as $0.99N^2$ is obtainable, which yields an angular
sensitivity $\Delta\beta\simeq1.005/N$. The insert of \cref{fig5}(b) indicates that the acceleration rate is outstanding in the small $N$ regime, while it starts to oscillate when $N$ increases, which, however, can be suppressed by slightly sacrificing the amount of achievable QFI as shown in \cref{fig5}(c). Taking $N=30$ as an example, we have the preparation time $t_f=0.28/\chi$ ($\approx 0.18 t_{\frac{\pi}{2}}$) while the achievable QFI is still as high as $0.97N^2$, which corresponds to a GHZ state that has a slight flaw as shown in the insert of \cref{fig5}(a).

\section{IMPLEMENTATIONS}
\label{sec:implementation}
Next, we show how to implement the cubic interaction in two-level atomic system.
\subsection{\label{sec:citeref}Cubic interaction induced by quadratic interactions}
In contrast to the cubic interaction, the quadratic interactions are much easier to implement in realistic atomic systems. Among them, the most widely studied one is the OAT interaction, which, as mentioned above, has been experimentally implemented in various atomic systems \cite{gross2010nonlinear,riedel2010atom,PhysRevL.114.4,PhysRevL.98.3,PhysRevL.105.8}. We next consider the realization of cubic interaction by repeated application of the OAT interactions. Suppose that one is able to freely apply the evolutions ${U_{k,q}}(\delta) = \exp \left( {i\delta{{S_q^k}}} \right)$ with $q \in \{ x,y,z\}$ to atomic system, for $k = 1,2$ and all $\delta \in \mathbb{R}$. The evolutions ${U_{1,q}}(\delta)$ denote a rotation of the collective spin around the $q$ axis by a phase $\delta$, which can be easily realized by applying either a RF magnetic field \cite{N1} or a circularly polarized optical pulse \cite{PhysRevLett.110.163602} to atoms along the $q$ direction. For the evolutions ${U_{2,q}}(\delta)$, it is only necessary to be able to perform the OAT evolution in a certain direction, e.g., ${U_{2,z}}(\delta)$. The rest OAT evolutions can be directly constructed from ${U_{1,k}}(\delta)$ and ${U_{2,z}}(\delta)$, such as ${U_{2,x}}(\delta)={U_{1,y}}(-\pi/2){U_{2,z}}(\delta){U_{1,y}}(\pi/2)$.
With these evolutions, other quadratic evolutions can then be constructed approximately by repeated application of ${U_{k,q}}(\delta)$ to atoms, e.g.,
\begin{eqnarray}
{E\left(S_x^2,S_y\right)} &=& {U_{2,x}}( \delta ){U_{1,y}}( \delta ){U_{2,x}}(-\delta ){U_{1,y}}(-\delta )\nonumber\\
 &=& {e^{ i\delta S_x^2}}{e^{ i\delta {S_y}}}{e^{-i\delta S_x^2}}{e^{-i\delta {S_y}}}\nonumber\\
 &=& {e^{[S_x^2,{S_y}]{\delta ^2}}} + O\left( {{\delta ^3}} \right)\nonumber\\
 &\approx& {e^{i{\delta ^2}\left({S_x}{S_z} + {S_z}{S_x}\right)}},\label{eq40}
\end{eqnarray}
in the limit $\delta\rightarrow 0$. Eq. (\ref{eq40}) is exactly the two-axis twisting evolution presented in Ref. \cite{PhysRevA.47.5138}. One can conclude from Eq. (\ref{eq40}) that the result of the transformation $E(A,B)$ is the same as if we have applied the interaction $i[A,B]$ to the atomic system for time $\delta^2$ \cite{PhysRevLett.82.1784}. With this in mind, let us consider the following transformations
\begin{eqnarray}
{U_C} &=& E\left( {{S_x},S_y^2} \right)E\left( {S_x^2,{S_y}} \right)E\left( {S_y^2,{S_x}} \right)E\left( {{S_y},S_x^2} \right)\nonumber\\
 &\approx& {e^{i{\delta ^2}\left( {{S_y}{S_z} + {S_z}{S_y}} \right)}}{e^{i{\delta ^2}\left( {{S_x}{S_z} + {S_z}{S_x}} \right)}}\nonumber\\
 &&\times {e^{ - i{\delta ^2}\left( {{S_y}{S_z} + {S_z}{S_y}} \right)}}{e^{ - i{\delta ^2}\left( {{S_x}{S_z} + {S_z}{S_x}} \right)}}\nonumber\\
 &=& E\left( {{S_y}{S_z} + {S_z}{S_y},{S_x}{S_z} + {S_z}{S_x}} \right)\nonumber\\
 &\approx& {e^{ - i8{\delta ^4}S_z^3 + i\left( {4S^2+4S - 1} \right){S_z}}},
\end{eqnarray}
where the first term is the desired cubic term and the second term linear in $S_z$ generates a precession of the collective spin around the $z$ axis. To isolate the cubic term it is convenient to apply a reverse-precession transformation to atoms, resulting in the overall effect
\begin{eqnarray}
{U_{1,z}}\left( {1 - 4S-4S^2} \right){U_C} \approx {e^{ - i8{\delta ^4}S_z^3}},
\end{eqnarray}
which is exactly the cubic evolution $U_z$ given in Eq. (\ref{eq5}) with $\chi t=8\delta^4$. We thus have successfully developed a general method, which should be widely applicable to a variety of spin systems capable of performing OAT evolution.
Finally, it should be pointed out that such a result has no counterpart in bosonic system, where linear together with quadratic Hamiltonians are unable to construct a Hamiltonian of higher order \cite{PhysRevLett.82.1784}.

\subsection{\label{sec:citeref}Cubic interaction induced by atoms-light interactions}
\begin{figure}[t]
	\centering
	\includegraphics[scale=0.7]{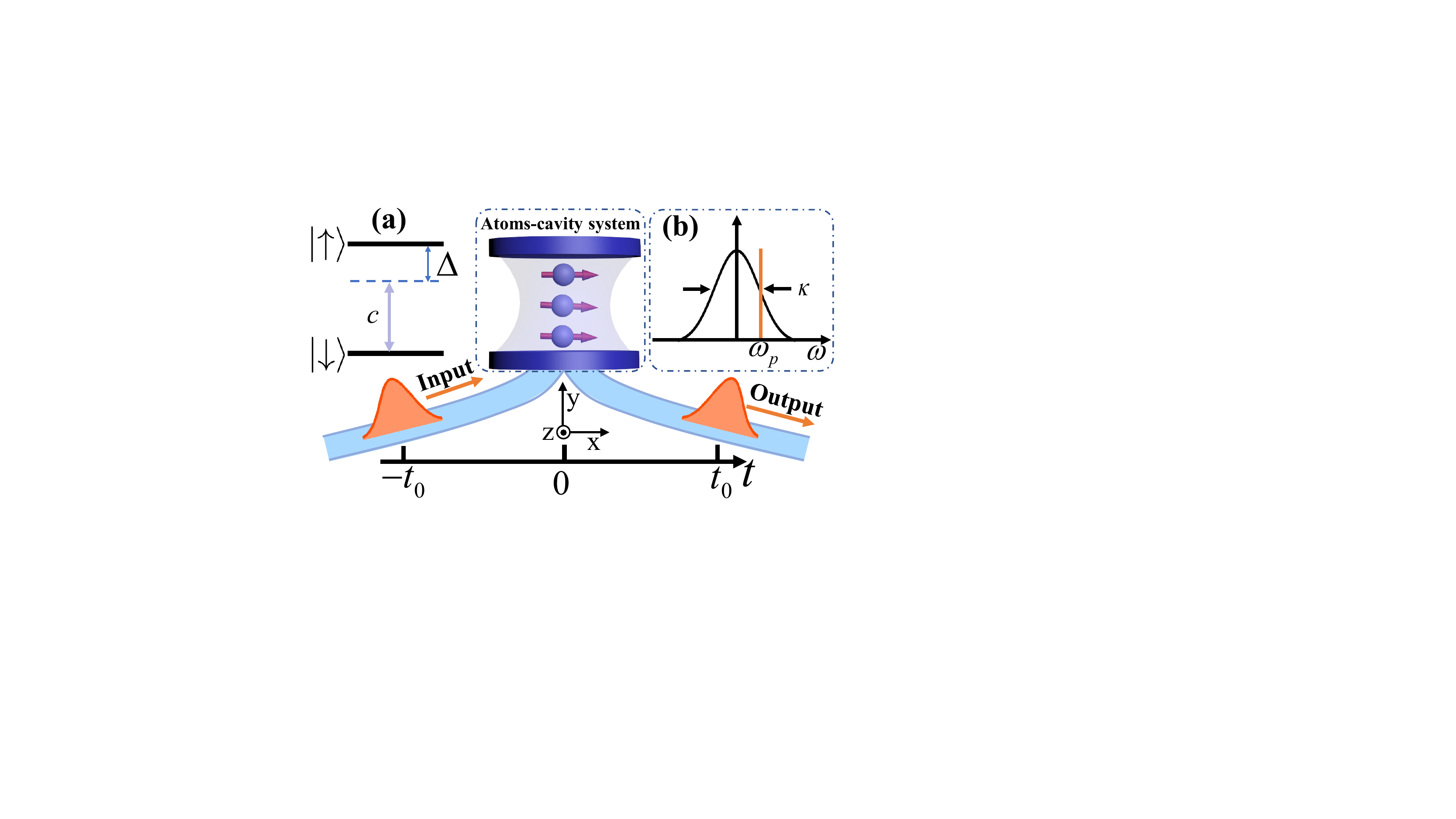}
	\caption{(Color online) Setup for realizing atomic cubic evolution. (a) Atoms with an excited state $\ket{\uparrow}$ and a ground state $\ket{\downarrow}$ couple off-resonantly to an one-sided optical cavity. At time $t=-t_0$ a light pulse is sent to interact with the cavity mode. After the light pulse is completely reflected by the cavity at $t=t_0$, a cubic-nonlinear transformation is successfully applied to the atoms. (b) The central frequency $\omega_p$ of the incident optical pulse is detuned from cavity resonance, with a detuning $\kappa/2$. }
\label{fig6}
\end{figure}

The above cubic method we developed relies on multi-step quantum operations, which might pose a technological challenge to a realistic implementation. We next show that it is possible to realize the cubic evolution in just \emph{one step} by utilizing properly designed light-mediated interactions.
We consider an ensemble of $N$ two-level atoms described above trapped inside a one-sided optical
cavity [see \cref{fig6}(a)]. The cavity field $c$ couples the two states $\ket{\downarrow}$ and $\ket{\uparrow}$ separated in energy by $\hbar\omega_a$ with a detuning $\Delta$ in the Tavis-Cummings model \cite{PhysRev.170.379}, which can be described by the Hamiltonian
\begin{eqnarray}
{H_{{\rm{cav}}}} = {\omega _a}{S_z} + {\left(\omega_a-\Delta\right)}{c^\dag }c + g{c^\dag }{S_ - } + {\rm{ H}}{\rm{.c}}{\rm{.}},\label{eq43}
\end{eqnarray}
where $g$ is the coupling constant. In the interaction picture with respect to $ {\omega _a}{S_z} + {\left(\omega_a-\Delta\right)}{c^\dag }c$, the Hamiltonian of Eq. (\ref{eq43}) can be expressed as:
\begin{eqnarray}
\tilde{H}_{\rm{cav}} = g{c^\dag }{S_ - }{e^{ - i\Delta t}} + {\rm{H}}{\rm{.c}}{\rm{.}}.
\end{eqnarray}
 We now consider the case that  $\Delta$ is much larger compared to $g$, to the linewidths of the cavity $\kappa$, and to the linewidths of atom $\Gamma$. Besides, we assume the intracavity photon number $\langle c^\dag c\rangle$ is very small. As a result, the population of the excited state is small,  which enables us to adiabatically eliminate the state $\ket{\uparrow}$ to
yield an effective Hamiltonian
\begin{eqnarray}
\tilde{H}_{\rm{cav}}^{eff} = \Omega\left( { 2{c^\dag }c{S_z}+{S_z}- S_z^2} \right)\label{eq45-2}
\end{eqnarray}
with $\Omega=g^2/\Delta$ \cite{epl}, where the first term denotes the ac Stark shift while the rest terms represent the cavity-mode-induced backaction of atoms onto themselves. The last two terms linear or quadratic in $S_z$ will cause a precession or a shearing of the atomic pseudospin around the $z$ axis, while the first term generates entanglement between the cavity mode and the collective spin, which in the past is an unwelcome term and usually decoupled to realize quadratic unitary transformation of atoms \cite{PhysRevLett.110.156402}. Instead, we next show that how it can be engineered to generate high-order nonlinear spin-spin dynamics. In the following analysis, we closely follow the procedure outlined in Ref. \cite{PhysRevA.85.013803}.

We assume that this cavity-atoms system is subjected to interact with an external field, which can be conveniently described by the input-output formalism \cite{PhysRevA.31.3761}, resulting in the effective Hamiltonian \cite{PhysRevA.85.013803}:
\begin{eqnarray}
H &=& \left( {{\omega _a} + \Omega} \right){S_z} +\left( {\omega _a-\Delta}\right){c^\dag }c + \int d \omega \omega b_\omega ^\dag {b_\omega }
 - \Omega S_z^2\nonumber\\ &&+ 2\Omega{S_z}{c^\dag }c + i\sqrt {\frac{\kappa }{{2\pi }}} \int d \omega \left( {b_\omega ^\dag c - {c^\dag }{b_\omega }} \right),
\end{eqnarray}
where the third term denotes the energy of the external field with the annihilation operators $b_\omega$ satisfying $[b_\omega,b_{\omega'}^\dag]=\delta(\omega-\omega')$, and the last term represents the coupling between the intracavity and
external fields through the partially transmissive input mirror of the cavity, and $\kappa$ stands for the cavity decay rate. Following the Fano's procedure \cite{PhysRev.124.1866}, this Hamiltonian can be exactly diagonalized to give
\begin{eqnarray}
H = \left( {{\omega _a} + \Omega} \right){S_z} - \Omega S_z^2 + \int d \omega \omega a_\omega ^\dag {a_\omega },\label{eq45}
\end{eqnarray}
where the dressed annihilation operators are
\begin{eqnarray}
{a_\omega } = {\alpha _\omega }c + \int {d\omega '{\beta _\omega }(\omega '){b_{\omega '}}}\label{eq46}
\end{eqnarray}
with
\begin{eqnarray}
\alpha_\omega &=& i\sin \left( {{\Delta _\omega }} \right)/\sqrt {\pi \kappa /2} ,\\
{\beta _\omega }(\omega ') &=& \frac{1}{\pi }{\cal P}\frac{\sin \left( {{\Delta _\omega }} \right)}{{\omega  - \omega '}} - \cos \left( {{\Delta _\omega }} \right)\delta (\omega  - \omega '),\label{eq48}\\
{\Delta _\omega } &=&  - \arctan \frac{{\kappa /2}}{{\omega  + \Delta  - {\omega _a} - 2\Omega {S_z}}},
\end{eqnarray}
where $\cal P$ denotes the principal part. The dressed operators $\{a_\omega\}$ obey the commutation relations $[a_\omega,a_{\omega'}^\dag]=\delta(\omega-\omega')$ and describe a set of decoupled harmonic oscillators. One can realize the photon excitations of $a_\omega$ by increasing the field amplitudes of either the intracavity mode or the outside continuum. A one-photon Fock state of $a_\omega$ can then be created by acting $a_\omega^\dag$ on the vacuum states, yielding
 \begin{eqnarray}
\left| {{1_\omega }} \right\rangle  &=& a_\omega ^\dag \left| {{0_c}} \right\rangle \otimes\left| {{0_{{b_\omega }}}} \right\rangle \nonumber\\
 &=& \alpha _\omega ^*\left| {{1_c}} \right\rangle \left| {{0_{{b_\omega }}}} \right\rangle  + \int {d\omega '{\beta^* _\omega }(\omega ')\left| {{1_{{b_{\omega '}}}}} \right\rangle } \left| {{0_c}} \right\rangle,\label{eq49}
\end{eqnarray}
which, for a given eigenstate $\ket{m}$ of $S_z$, is the eigenstate of $H$, that is, $H\ket{1_\omega}\otimes\ket{m}=[(\omega_a+\Omega) m-\Omega^2 m^2+\omega]\ket{1_\omega}\ket{m}$.

Now, with the help of Eq. (\ref{eq45}) we are able to describe the interaction between a light pulse and the atoms-cavity system. For simplicity, we consider the case that a single-photon pulse is sent to the cavity at the initial time $t=-t_0$ that is far in the past, which has the form $\left| {{1_b}} \right\rangle  = \int d \omega {e^{i\omega {t_0}}}B(\omega )|1_{b_{\omega}}\rangle$, where $B(\omega)$  is
the normalized probability amplitude as a function of frequency, and the cavity field starts in the vacuum, leading to the initial state of the fields, $\ket{\psi_f}_b=\ket{1_b}\otimes\ket{0_c}$. One thus can express the initial state of the system as
\begin{eqnarray}
\left| {\Psi \left( { - {t_0}} \right)} \right\rangle  &=& \left| {{\psi _a}} \right\rangle  \otimes {\left| {{\psi _f}} \right\rangle _b}\nonumber\\
 &=& \sum\limits_{m =  - S}^S {{C_m}\int {d\omega \int {d\omega '{\beta _\omega }(\omega ') } } } \nonumber\\&&\times{e^{i\omega '{t_0}}}B(\omega ')\left| {{1_\omega }} \right\rangle|m\rangle\nonumber\\
 &=&  - \sum\limits_{m =  - S}^S {{C_m}\int {d\omega {e^{i\left( {\omega{t_0}  + {\Delta _\omega }} \right)}}B(\omega )\left| {{1_\omega }} \right\rangle|m\rangle } },\label{eq51}\nonumber\\
\end{eqnarray}
\begin{figure*}[bt]
	\centering
	\includegraphics[scale=0.8]{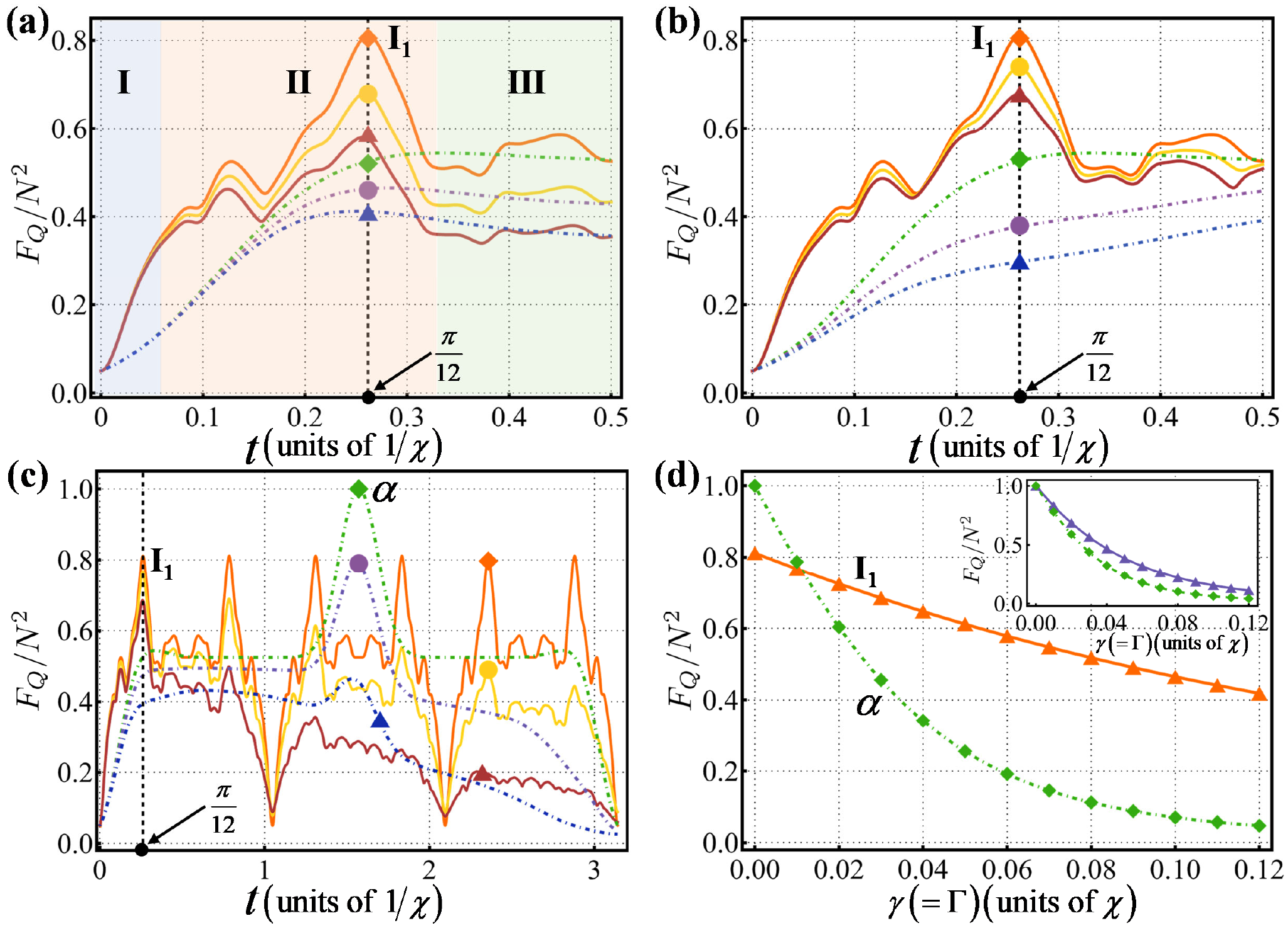}
	\caption{(Color online) Performance of the cubic scheme (solid curves) and the OAT scheme (dot-dashed curves) in the presence of damping (in units of $\chi$) for $N=20$. QFI created by the two schemes versus interaction time (in units of $1/\chi$), in the presence of either single-spin decay (a) or collective dephasing (b): no loss (diamond), $5\%$ loss (circle), $10\%$ loss (triangle). (c) QFI created by the two schemes versus interaction time, in the presence of both decays: no loss (diamond), $\gamma=\Gamma=5\%$ (circle), $\gamma=\Gamma=10\%$ (triangle). (d) QFI of the states of the peaks marked $\text{I}_1$ and $\alpha$ in (c) as a function of $\gamma(=\Gamma)$. Insert: QFI of the GHZ state created by the cubic scheme (line with triangles) and the OAT scheme (line with diamonds) as a function of $\gamma(=\Gamma)$ for $N=21$. }
\label{fig7}
\end{figure*}
where $C_m$ is the normalized probability amplitude of an arbitrary atomic state $\ket{\psi_a}$ and in the second equality we also have reexpressed the initial state $\ket{\psi_f}_b$ in terms of the eigenstates of the total field by using Eq. (\ref{eq49}). The phase $\Delta_\omega$ in the last equality can be understood as a phase lag of the cavity field response to the drive of the outside continuum photon at frequency $\omega$ \cite{PhysRevA.85.013803}. The initial state (\ref{eq51}), under the action of the unitary evolution
generated by Eq. (\ref{eq45}) for time $2t_0$ that is long enough to allow the light pulse to be completely reflected from the cavity, transforms to
\begin{eqnarray}
\left| {\Psi \left( {{t_0}} \right)} \right\rangle  &=& \exp \left( { - iH{2t_0}} \right)\left| {\Psi \left( { - {t_0}} \right)} \right\rangle \nonumber\\
 &=& \sum\limits_{m =  - S}^S {{C_m}{e^{ - i2{t_0}\left[ {\left( {{\omega _a} + \Omega } \right)m - \Omega {m^2}} \right]}}} \nonumber\\
 &&\times \int {d\omega {e^{i\left( {2{\Delta _\omega } - \omega{t_0} } \right)}}B(\omega )\left| {{1_{{b_\omega }}}} \right\rangle \left| m \right\rangle }\label{eq52},
\end{eqnarray}
where the field states have been transformed back into $\ket{1_{b_\omega}}$. Obviously, the output pulse get entangled with atoms as the phase factor $\Delta_w$ in Eq. (\ref{eq52}) contains information about the atoms. However, if the input light pulse is near-monochromatic with frequency $\omega_p$ and its bandwidth is much less than the linewidth of cavity [see \cref{fig6}(b)], then one may make the approximation $\Delta_w\rightarrow \Delta_{w_p}$ \cite{PhysRevA.85.013803}, and the output state (\ref{eq52}) becomes
\begin{eqnarray}
\left| {\Psi \left( {{t_0}} \right)} \right\rangle  &=& \sum\limits_{m =  - S}^S {{C_m}{e^{ - 2i\left[ {\left( {{\omega _a} + \Omega } \right){t_0}m - \Omega{t_0} {m^2} - {\Delta _{{\omega _p}}}} \right]}}} \left| m \right\rangle \nonumber\\
 &&\otimes \int {d\omega {e^{ - i\omega {t_0}}}B(\omega )\left| {{1_{{b_\omega }}}} \right\rangle },
\end{eqnarray}
which shows the output pulse is completely disentangled from the atoms. Although the interaction process has nothing to do
with the incident photon, it imposes an additional $m$-dependent phase shift $2\Delta_{\omega_p}$ to atoms, resulting in the transformation of atomic state by the unitary operator
\begin{eqnarray}
U = {e^{ - 2i\left[ {\left( {{\omega _a} + \Omega } \right){t_0}{S_z} - \Omega {t_0}S_z^2 - {\Delta _{{\omega _p}}}} \right]}},
\end{eqnarray}
indicating that the atomic subspace has experienced a linear- and quadratic-$S_z$ (OAT) interactions, while the higher order interactions of interest are encoded in $\Delta_{\omega_p}$. Next, we assume that the incident photon is off-resonance with the cavity mode with frequency $\omega_p=\omega_a-\Delta+\kappa/2$ and the cavity resonance is not shifted too much by atoms \cite{PhysRevA.81.021804}, such that $\kappa_0=\Omega/\kappa\ll 1/N$, the operator $\Delta_{\omega_p}$ can then be expanded up to third order in the parameter $\kappa_0$, generating
\begin{eqnarray}
U &=& {e^{ - 2i\left\{ {\left[ {\left( {{\omega _a} + \Omega } \right){t_0} + 2{\kappa _0}} \right]{S_z} + \left( {4\kappa _0^2 - \Omega {t_0}} \right)S_z^2 + \frac{{16}}{3}\kappa _0^3S_z^3} \right\}}}\nonumber\\
 &=& {e^{ - 2i\left\{ {\left[ {\left( {{\omega _a} + \Omega } \right){t_0} + 2{\kappa _0}} \right]{S_z} + \frac{{16}}{3}\kappa _0^3S_z^3} \right\}}},
\end{eqnarray}
where in the second equality we have set ${{t_0} = 4\kappa _0^2/\Omega }$. Analogously,  the cubic term can be isolated by applying a linear counter-rotating transformation around $z$ axis to atoms, finally arriving at
\begin{eqnarray}
U_1={U_{1,z}}\left[ {2\left( {{\omega _a} + \Omega } \right){t_0} + 4{\kappa _0}} \right]U = {e^{ - i\frac{{32}}{3}\kappa _0^3S_z^3}}.\label{eq56}
\end{eqnarray}
Thus, we have successfully realized the cubic evolution of the atomic state by simply injecting a single-photon state into an atoms-cavity system.

A major concern of our proposed scheme might be the coupling constant $\kappa_0$, as it turns to be extremely weak when $N$ is large. One direct way to enhance the coupling strength is to use more incident single photons. Suppose that we have $n$ single-photon wave packets described above, and they are sent one by one to interact with the cavity mode at fixed interval $2t_0$. Each photon induces a  $U_1$ transformation to atoms, then, after the $n$th interaction, the atomic state evolves as if it has been applied a unitary transformation
\begin{eqnarray}
{U_n} = {e^{ - i{\mu _n}S_z^3}}\label{eq57}
\end{eqnarray}
with $\mu_n=32n\kappa_0^3/3=n(\eta\Gamma/\Delta)^3/6$, where $\eta=4g^2/(\kappa\Gamma)$ is the single-atom cooperativity. Eq. (\ref{eq57}) indicates that the total coupling strength is now $n$ times lager than the coupling strength created by a single photon. Another convenient way is to utilize a $n$-Fock state. If the incident wave packet contains exactly $n$ photons, its interaction with cavity system would also lead to the cubic evolution (\ref{eq57}) \cite{PhysRevA.85.013803}. As a specific example, for an atomic ensemble with atom number $N\sim 10^3$, if we take $\eta=0.04$, $\Gamma=10g$, and $\Delta=150g$, a coupling constant $\alpha\sim 3$ is obtainable with the choice $n\sim 10^6$, which enables the production of QFI as high as $F_{Q}\sim N^2/2$ [according to Eq. (\ref{eq16})] with the interaction time $2t_0\sim $1 ns for $g/(2\pi)\sim 1$ Mhz.
%$\Gamma=\kappa=10g$

\section{The effect of damping}
\label{sec:noise}

Up till now we have only considered the perfect evolution of the spin state. In realistic systems, however, there are inevitable noise effect that will cause damping of the spin state. Here, we mainly consider two types of damping: one arises because of external field fluctuations \cite{PhysRevA.64.052106,PhysRevA.80.023609,PhysRevA.82.045601}, which induces collective dephasing of the atoms, and its influence on the evolution of the atomic state $\rho$ can be described by the dissipative superoperator $\mathcal{D}[S_z]\rho$, where $ {\cal D}[O]\rho  = 2O\rho {O^\dag } - \{ {O^\dag }O,\rho \}$ is  the standard Lindblad dissipative superoperator; another one is single-spin decay, which is normally caused by the spontaneous emission of photons by the individual atoms into free space \cite{PhysRevLett.110.120402,PhysRevA.96.050301} and can be described by the dissipative superoperator $\sum_k\mathcal{D}[\sigma_-^k]\rho$, where $\sigma_-^k$ is the pseudo-spin lowering operator for the $k$th atom. With these damping, the master equation
for the atomic state $\rho$ under the interaction of $H$ can be expressed as
\begin{eqnarray}
\dot \rho  =  - i[H,\rho ] + \Gamma {\cal D}[S_z]\hat \rho  + \gamma\sum\limits_k {\cal D} \left[ {\sigma _-^k } \right]\rho,\label{eq58}
\end{eqnarray}
where $\gamma$ is the decay rate of the excited state $\ket{\uparrow}$ and $\Gamma$ is the collective dephasing rate.

The numerical solutions of Eq. (\ref{eq58}) are shown in \cref{fig7} for both $H=\chi S_z^2$ and $H=\chi S_z^3$. In \cref{fig7}(a)
we plot the QFI evolves with time $t$ (in units of $1/\chi$) for $\Gamma=0$ (in units of $\chi$) and various values of $\gamma$ (in units of $\chi$). In the weak coupling regime (region I), the dampings have only small impact on the processes of entanglement creation for both protocols. Thus, in this region our cubic scheme can still maintain its speed advantage in QFI creation. As the QFI increases (region II), the cubic scheme is more susceptible to single-spin decay. However, benefiting from the advantage of quantity, the QFI of the cubic scheme before time $t=\pi/12$ are still larger the QFI created by the ideal OAT evolution, even when $\gamma=0.1$. In the region labelled III, the noise effect almost has the same impact on the two protocols in the entanglement generation.
Figure 7(b) plots the QFI in its dependence on $t$ for $\gamma=0$ and various values of $\Gamma$, which indicates that our cubic method is considerably more \emph{robust against} the collective dephasing as compared to the OAT method. In the presence of both single- and collective-spin decay, we also plot the achievable QFI versus $t$ for both scheme in \cref{fig7}(c), showing that the Heisenberg-limited QFI peak $\alpha$ (produced by the OAT evolution at time $t_{\frac{\pi}{2}}$) decreases rapidly with the atomic decay. Although, in the ideal case, the peak $I_1$ (produced by the cubic evolution at time $t_{\frac{\pi}{12}}$) has the disadvantage of less QFI (in contrast to the peak $\alpha$), it is much more robust against decoherence [as also compared in Fig. 7(d)], which makes it more attractive in a realistic implementation. For odd $N$, a comparison of the QFI of the GHZ state created by both protocols [see the insert of Fig. 7(d)] indicates that the GHZ created by our cubic scheme is less susceptible to decoherence.

\section{CONCLUSIONS}
\label{sec:conclution}
In this paper, we have proposed to entangle individual spins using the cubic nonlinear interaction. We find that, although the multipartite entangled states created by the cubic scheme have no spin squeezing, they are useful for quantum metrology. In contrast to the traditional OAT scheme, we have shown that the cubic scheme offers several advantages. First, it can produce QFI much more rapidly in the weak coupling regime. The larger the total spin number $N$, the larger the acceleration rate, which makes it particularly attractive for entanglement generation in large-number spin system.
Second, the cubic scheme enables the preparation of a broad variety of new-type macroscopic superposition states in a much more short time. We showed that these states exhibit an outstanding performance in the generation of large QFI, which provide the possibility to realize near-HL phase sensitivity. Third, the cubic scheme is still capable of producing much more spin-spin entanglement even in the presence of large decays.

We also discovered a new even-odd effect of the cubic evolution, that is,  the entanglement created by the cubic evolution is extremely macroscopic
sensitive to the parity of the total spin number $N$. We showed that such entanglement even-odd effect might be exploited to design new type of sensor modality, enabling the determination of the parity of the spin number in a spin system at the single-spin level. We also find a new mechanism to generate high-fidelity GHZ states. By using a hybrid NSS interaction---CQA type of nonlinear interaction, one may speed up the preparation of GHZ states as compared to the methods based solely upon OAT interaction.

We also have presented two approaches to realize the cubic evolution of the spin system. One is based on the lower-order interactions. We showed that the cubic evolution can be approximately constructed by repeatedly using linear- and quadratic-nonlinear dynamics. This method is quite general and is widely applicable to a variety of spin systems. Another one relies on the light-mediated interactions. We found that, by suitably engineering the light-mediated interactions, one is able to realize the cubic NSS interaction among atoms in just one step.

Our study provides a new angle in utilizing unitary transformation to produce \emph{useful} multipartite entanglement among spins. Although the cubic (third-order) nonlinearity in a realistic spin system is normally weak, it would greatly enrich the way of manipulating the spin system, just like the Kerr nonlinearity in optical system \cite{PhysRevLett.91.093601,PhysRevLett.93.083904}. We thus believe that the application of the cubic interaction will not be restricted to the field of entanglement generation. For instance, it has been shown that the bosonized spin system is an excellent platform for implementing continuous-variables quantum information processing \cite{Book1}. The cubic evolution (called the cubic phase gate) then is a particularly convenient candidate for realizing non-Gaussian operations \cite{PhysRevLett.124.240503}. Therefore, our proposed schemes would also benefit the field of continuous-variables quantum computation.

\begin{acknowledgments}
We thank Yanhong Xiao for helpful discussions. This work was supported by the Natural Science Foundation
of China (Grants No. 22273067), the Natural Science Foundation of Zhejiang province, China (Grant No. LQ23A040001), and the Department of Education of Zhejiang Province, China (Grant No. Y202146469).
\end{acknowledgments}

\appendix
	\section{Converting the binomial distribution into the Gaussian Distribution}
\label{App:a}
In this Appendix, we give the details of the derivation of Eq.~(\ref{eq12}). Defining $M=2S-1$, Eq.~(\ref{eq12}) can be rewritten as
\begin{equation}
	\begin{aligned}
		P_M(m) = \frac{{M!}}{{\left( {M - m} \right)!}m!}{2^{-M}}.	
	\end{aligned}\label{A1}
\end{equation}	
As shown in the main text, this binomial distribution has the mean $\langle m\rangle=S$ and standard deviation $\Delta m\sim \sqrt{S}$. We thus can use the Stirling's approximation $x! \approx \sqrt {2\pi x} {\left( {{x}/{e}} \right)^x}$ for large $S$, obtaining
\begin{eqnarray}
{P_M}(m) &=& \frac{{\sqrt {2\pi M} {M^M}{2^{ - M}}}}{{\sqrt {2\pi \left( {M - m} \right)} {{\left( {M - m} \right)}^{M - m}}\sqrt {2\pi m} {M^m}}}\nonumber\\
 &=& \frac{1}{{\sqrt {2M\pi } }}{\left( {2 - 2\frac{m}{M}} \right)^{ - M}}{\left( {\frac{{m/M}}{{1 - m/M}}} \right)^{ - m}}\nonumber\\
 &&\times {\left( {\frac{m}{M} - \frac{{{m^2}}}{{{M^2}}}} \right)^{ - 1/2}}.\label{A2_2}
\end{eqnarray}
Taking the logarithm of Eq.~(\ref{A2_2}) leads to
\begin{eqnarray}
\ln {P_M}(m) &=& \ln \frac{1}{{\sqrt {2M\pi } }} - M\ln \left( {2 - 2\frac{m}{M}} \right)\nonumber\\
 &&- m\ln \left( {\frac{{m/M}}{{1 - m/M}}} \right) - \frac{1}{2}\ln \left( {\frac{m}{M} - \frac{{{m^2}}}{{{M^2}}}} \right).\nonumber\\
\end{eqnarray}
Using Taylor expansion around $\langle m\rangle=M/2$, we arrive at
\begin{eqnarray}
\ln {P_M}(m) &\simeq& \ln \frac{2}{{\sqrt {2M\pi } }} + 2\left( {m - \frac{M}{2}} \right)\nonumber\\
 &&+ \frac{2}{M}{\left( {m - \frac{M}{2}} \right)^2} - 2\left( {m - \frac{M}{2}} \right)\nonumber\\
 &&- \frac{4}{M}{\left( {m - \frac{M}{2}} \right)^2} + \frac{2}{{{M^2}}}{\left( {m - \frac{M}{2}} \right)^2}\nonumber\\
 &\simeq& \ln \frac{2}{{\sqrt {2M\pi } }} - \frac{2}{M}{\left( {m - \frac{M}{2}} \right)^2},\label{A4_4}
\end{eqnarray}
where we have kept terms to second order and omitted the last term in the first equality as it is much more smaller when $M$ is large. Finally, we exponentiate Eq. (\ref{A4_4}) to get
\begin{eqnarray}
{P_M}(m) &=& \frac{2}{{\sqrt {2M\pi } }}\exp \left[- {\frac{{2{{\left( {m - \frac{M}{2}} \right)}^2}}}{M}} \right]\nonumber\\
 &\simeq& \frac{1}{{\sqrt {S\pi } }}\exp \left[- {\frac{{{{\left( {m - S} \right)}^2}}}{S}} \right],
\end{eqnarray}
where the last equality is valid for large $S$.

\section{Calculation of the expectation values of the evolved spin operators }
\label{appb}
To calculate the means and variances of Eqs. (\ref{eq14}), one needs to calculate the expectation values $\left\langle {{S_ + }\left( t \right)} \right\rangle $, $\left\langle {S_ + ^2\left( t \right)} \right\rangle $, and $\left\langle {{S_ + }\left( t \right){S_ - }\left( t \right)} \right\rangle$.
Along the same line as calculating $\left\langle {{S_ + }\left( t \right)} \right\rangle$ given in the Eq. (\ref{eq11}), one may derive the quadratic expectation value
\begin{eqnarray}
			\left\langle {S_ + ^2\left( t \right)} \right\rangle  &=& {2^{ - 2S}}\sum\limits_{k = 0}^{2S} {\sum\limits_{l = 0}^{2S} {\sqrt {\frac{{(2S)!}}{{(2S - k)!k!}}} \sqrt {\frac{{(2S)!}}{{(2S - l)!l!}}} } } \nonumber\\
		&&\times \left\langle {S,S - k} \right|{\left[ {{S_ + }{{\rm{e}}^{i\mu \left( {S_z^2 + {S_z} + {1 \mathord{\left/
									{\vphantom {1 3}} \right.
									\kern-\nulldelimiterspace} 3}} \right)}}} \right]^2}\left| {S,S - l} \right\rangle \nonumber\\
		&=& {2^{ - 2S}}\sum\limits_{l = 2}^{2S} {\frac{{(2S)!}}{{(2S - l)!(l - 2)!}}} {{\rm{e}}^{2i\mu \left[ {{{\left( {S - l + 1} \right)}^2} + \frac{1}{3}} \right]}}\nonumber\\
		&=&S\left( {S - \frac{1}{2}} \right)\sum\limits_{m = 0}^{2S - 2} {{P_{2S - 2}}\left( m \right)} \nonumber\\
		&\times& {{\rm{e}}^{2i\mu \left[ {{{\left( {S - m} \right)}^2} - 2\left( {S - m} \right) + \frac{4}{3}} \right]}},\label{B2}
\end{eqnarray}
where we set $m=l-2$. Again, the binomial distribution ${P_{2S - 2}}\left( m \right)$ can be transformed into the Gaussian distribution to give
\begin{eqnarray}
			\left\langle {S_ + ^2\left( t \right)} \right\rangle  &\simeq& \frac{{S\left( {S - {1 \mathord{\left/
							{\vphantom {1 2}} \right.
							\kern-\nulldelimiterspace} 2}} \right)}}{{\sqrt {\pi S}}}\sum\limits_{k =  - \sqrt S }^{\sqrt S } {{e^{ - \left( {1 - 2i\mu S} \right){k^2} - 4i\mu \sqrt S k + \frac{8}{3}i\mu }}} \nonumber\\
		&\simeq&\frac{{S\left( {S - {1 \mathord{\left/
							{\vphantom {1 2}} \right.
							\kern-\nulldelimiterspace} 2}} \right)}}{{\sqrt \pi  }}{e^{\frac{8}{3}i\mu }}\nonumber\\
		&&\times \int_{ - \infty }^{ + \infty } {{e^{ - \left( {1 - 2i\mu S} \right){k^2} - 4i\mu \sqrt S k}}} dk\nonumber\\
&\simeq& \frac{{S\left( {S - {1 \mathord{\left/
							{\vphantom {1 2}} \right.
							\kern-\nulldelimiterspace} 2}} \right)}}{{\sqrt { {1 - 2i\mu S}} }},\label{B4}
\end{eqnarray}
and its complex conjugate
\begin{eqnarray}
			\left\langle {S_ - ^2\left( t \right)} \right\rangle  \simeq \frac{{S\left( {S - {1 \mathord{\left/
							{\vphantom {1 2}} \right.
							\kern-\nulldelimiterspace} 2}} \right)}}{{\sqrt { {1 + 2i\mu S} } }}.	\label{B6}
\end{eqnarray}
For the mean $\left\langle {{S_ + }\left( t \right){S_ - }\left( t \right)} \right\rangle $, it can be directly calculated without approximation as
\begin{eqnarray}
			\left\langle {{S_ + }\left( t \right){S_ - }\left( t \right)} \right\rangle &=&\left\langle {{S_ - }\left( t \right){S_ + }\left( t \right)} \right\rangle\nonumber\\&=& {2^{ - 2S}}\sum\limits_{k = 0}^{2S} {\sum\limits_{l = 0}^{2S} {\sqrt {\frac{{\left( {2S} \right)!}}{{\left( {2S - k} \right)!k!}}} \sqrt {\frac{{\left( {2S} \right)!}}{{\left( {2S - l} \right)!l!}}} } } \nonumber\\
		&& \times \left\langle {S,S - k} \right|{S_ + }{S_ - }\left| {S,S - l} \right\rangle \nonumber\\
		&=&{2^{ - 2S}}\sum\limits_{l = 0}^{2S} {\frac{{\left( {2S} \right)!}}{{\left( {2S - l - 1} \right)!l!}}} \left( {l + 1} \right)\nonumber\\
		&=& {S^2}+\frac{S}{2} .\label{B7}
\end{eqnarray}
Using Eqs. (\ref{B4})-(\ref{B7}) and (\ref{eq13}), we are able to calculate the means and variances of Eqs. (\ref{eq14}). Here, we derive the mean $\left\langle {{S_x}\left( t \right)} \right\rangle $ as an example.  According to Eq. (\ref{eq13}) we have
\begin{eqnarray}
		\left\langle {{S_x}\left( t \right)} \right\rangle &=& \frac{1}{2}\left( {\left\langle {{S_ + }\left( t \right)} \right\rangle  + \left\langle {{S_ - }\left( t \right)} \right\rangle } \right)\nonumber\\
		&=&\frac{S}{2}\frac{{\left( {\sqrt {1 + i\mu S} {\rm{ + }}\sqrt {1 - i\mu S} } \right)}}{{\sqrt {1{\rm{ + }}{\mu ^2}{S^2}} }}.\label{B9_9}
\end{eqnarray}
Next, defining $\alpha_0 = \arccos(1/\sqrt {1{\rm{ + }}{\mu ^2}{S^2}}) $, Eq. (\ref{B9_9}) can be reexpressed as
\begin{eqnarray}
\left\langle {{S_x}\left( t \right)} \right\rangle  &=& \frac{S}{{\sqrt[4]{{1{\rm{ + }}{\mu ^2}{S^2}}}}}\cos \frac{{{\alpha _0}}}{2}\nonumber\\
 &=& S\sqrt {\frac{{1 + {{\left( {1{\rm{ + }}{\mu ^2}{S^2}} \right)}^{1/2}}}}{{2(1{\rm{ + }}{\mu ^2}{S^2})}}} \nonumber\\
 &=& S\sqrt {\alpha_1 \left( {\alpha_1  + 1} \right)/2},
\end{eqnarray}
 where $\alpha_1  = 1/\sqrt {1{\rm{ + }}{\mu ^2}{S^2}} $.

\section{Calculation of the expectation values of the nondiagonal elements}
\label{appc}
In this Appendix, we prove that it is reasonable to ignore the nondiagonal terms in the calculation of Eq. (\ref{eq31}).
Suppose that $\ket{\frac{\pi}{2},\phi_1}$ and $\ket{\frac{\pi}{2},\phi_2}$ with $\phi_1\neq\phi_2$ are two CSS components encoded in the spin states of Eq. (\ref{eq32}). Corresponding to this two CSSs, one needs to evaluate the nondiagonal values $\bra{\frac{\pi}{2},\phi_1}S_\phi\ket{\frac{\pi}{2},\phi_2}$ and $\bra{\frac{\pi}{2},\phi_1}S_\phi^2\ket{\frac{\pi}{2},\phi_2}$ when deriving the projection noises of Eq. (\ref{eq33}). Let us first derive
\begin{widetext}
\begin{eqnarray}
\left\langle {\frac{\pi }{2},{\phi _1}} \right|{S_\phi }\left| {\frac{\pi }{2},{\phi _2}} \right\rangle  &=& {\frac{1}{2}e^{ - i\phi }}\left\langle {\frac{\pi }{2},{\phi _1}} \right|{S_ + }\left| {\frac{\pi }{2},{\phi _2}} \right\rangle + {\frac{1}{2}e^{i\phi }}\left\langle {\frac{\pi }{2},{\phi _1}} \right|{S_ - }\left| {\frac{\pi }{2},{\phi _2}} \right\rangle \nonumber\\&=& \frac{S}{2}\cos {\left( {\frac{{\Delta \phi }}{2}} \right)^{2S - 1}}\cos \left( {\frac{{\Delta \phi }}{2} + {\phi _1}-\phi} \right){{\rm{e}}^{iS\Delta \phi }},\label{C1_1}
\end{eqnarray}
\end{widetext}
where we have defined the new parameter $\Delta \phi {\rm{  =  }}{\phi _2} - {\phi _1}$. Obviously, for a nonzero $\Delta\phi$, the value of Eq. (\ref{C1_1}) decays exponentially with $S$, resulting in $\bra{\frac{\pi}{2},\phi_1}S_\phi\ket{\frac{\pi}{2},\phi_2}\approx 0$ for the case of large $S$.

We now turn to evaluate the second moments
\begin{widetext}
\begin{eqnarray}
\left\langle {\frac{\pi }{2},{\phi _1}} \right|S_\phi ^2\left| {\frac{\pi }{2},{\phi _2}} \right\rangle  &=& {\frac{1}{4}e^{ - 2i\phi }}\left\langle {\frac{\pi }{2},{\phi _1}} \right|S_ + ^2\left| {\frac{\pi }{2},{\phi _2}} \right\rangle
+ {\frac{1}{4}e^{2i\phi }}\left\langle {\frac{\pi }{2},{\phi _1}} \right|S_ - ^2\left| {\frac{\pi }{2},{\phi _2}} \right\rangle
\nonumber\\&&+ \frac{1}{4}\left\langle {\frac{\pi }{2},{\phi _1}} \right|{S_ + }{S_ - }\left| {\frac{\pi }{2},{\phi _2}} \right\rangle
 + \frac{1}{4}\left\langle {\frac{\pi }{2},{\phi _1}} \right|{S_ - }{S_ + }\left| {\frac{\pi }{2},{\phi _2}} \right\rangle.\label{C2}
\end{eqnarray}
\end{widetext}
To derive Eq. (\ref{C2}), we need to calculate the element
\begin{widetext}
\begin{eqnarray}
			\left\langle {\frac{\pi }{2},{\phi _1}} \right|S_ + ^2\left| {\frac{\pi }{2},{\phi _2}} \right\rangle  &=& {2^{ - \left( {2S + 1} \right)}}\sum\limits_{l = 0}^{2S} {\sum\limits_{k = 0}^{2S} {\sqrt {\frac{{2S!}}{{(2S - k)!k!}}} }  } \sqrt {\frac{{2S!}}{{(2S - l)!l!}}} {{\mathop{\rm e}\nolimits} ^{ik\Delta \phi  + 2i{\phi _1}}} \left\langle {S - l} \right|S_ + ^2\left| {S - k} \right\rangle\nonumber\\
		&=& {2^{ - \left( {2S + 1} \right)}}\sum\limits_{k = 0}^{2S} {\frac{{\left( {2S} \right)!}}{{(2S - k)!(k - 2)!}}} {{\mathop{\rm e}\nolimits} ^{ik\Delta \phi  + 2i{\phi _1}}}\nonumber\\
 &=&\frac{1}{4}{S\left( {2S - 1} \right)}\cos {\left( {\frac{{\Delta \phi }}{2}} \right)^{2S - 2}}{{\mathop{\rm e}\nolimits} ^{i {\Delta \phi (S+1)}+ 2i{\phi _1}}}.\label{C3}
\end{eqnarray}
\end{widetext}
Analogously, one may derive the elements
\begin{widetext}
\begin{eqnarray}
\left\langle {\frac{\pi }{2},{\phi _1}} \right|S_ - ^2\left| {\frac{\pi }{2},{\phi _2}} \right\rangle  &=& \frac{1}{4}{S\left( {2S - 1} \right)}\cos {\left( {\frac{{\Delta \phi }}{2}} \right)^{2S - 2}} {{\rm{e}}^{ i\Delta \phi \left( {S - 1} \right) - 2i{\phi _1}}},\label{C4}\\
\left\langle {\frac{\pi }{2},{\phi _1}} \right|{S_ + }{S_ - }\left| {\frac{\pi }{2},{\phi _2}} \right\rangle  &=& \left\langle {\frac{\pi }{2},{\phi _1}} \right|{S_ - }{S_ + }\left| {\frac{\pi }{2},{\phi _2}} \right\rangle \nonumber\\
 &=& \cos {\left( {\frac{{\Delta \phi }}{2}} \right)^{2S - 2}}\left[ {\frac{S}{2}{{\rm{e}}^{ - i\frac{{\Delta \phi }}{2}}}} \right.\left. { \cos \left( {\frac{{\Delta \phi }}{2}} \right)+ \frac{{S\left( {2S - 1} \right)}}{4}} \right]{{\rm{e}}^{i\Delta \phi S}}.\label{C5}
\end{eqnarray}
\end{widetext}
Obviously, the values of Eqs. (\ref{C3})-(\ref{C5}) also tend to zero for large $S$. We thus can conclude that the nondiagonal terms in Eq. (\ref{eq31}) can be neglected when $N$ is large.

\newpage

\nocite{*}

\bibliography{aref}% Produces the bibliography via BibTeX.

\end{document}